\documentclass[aps,prd,longbibliography,nofootinbib]{revtex4-2}

\usepackage{graphicx}
\usepackage{amsmath,amssymb,bm}
\usepackage{siunitx}
\DeclareSIUnit\year{yr}
\DeclareSIUnit\liter{L}

\usepackage{comment}
\usepackage{booktabs}
\usepackage{microtype}
\usepackage{natbib}
\usepackage{hyperref}
\hypersetup{colorlinks=true,allcolors=blue}

\usepackage{array}

\usepackage{tikz}
\usepackage{pgfplots}
\pgfplotsset{
compat=1.17,
every axis/.append style={
line width=0.9pt,
tick style={line width=0.8pt},
tick label style={font=\small},
label style={font=\small},
legend style={font=\small, fill=white, draw=black},
grid=major,
grid style={gray!25},
major grid style={gray!35},
legend cell align=left
}
}

\usepackage{xparse}

\newcommand{\Pitn}{\ensuremath{P_{\mathrm{IT},n}}}

\newcommand{\Pavg}{\ensuremath{P_{\mathrm{avg}}}}
\newcommand{\Pname}{\ensuremath{P_{\mathrm{name}}}}
\newcommand{\Util}{\ensuremath{U}}
\newcommand{\Eyr}{\ensuremath{E_{\mathrm{yr}}}}

\newcommand{\Cterr}{\ensuremath{C_{\mathrm{terr}}}} 
\newcommand{\Clink}{\ensuremath{C_{\mathrm{link}}}} 
\newcommand{\CGS}{\Clink}                            
\newcommand{\Corbit}{\ensuremath{C_{\mathrm{orbit}}}}
\newcommand{\mkwfixed}{\ensuremath{m_{\mathrm{kW},\mathrm{fixed}}}}
\newcommand{\mkwPV}{\ensuremath{m_{\mathrm{kW},\mathrm{PV}}}}
\newcommand{\mkwrad}{\ensuremath{m_{\mathrm{kW},\mathrm{rad}}}}
\newcommand{\mkwbat}{\ensuremath{m_{\mathrm{kW},\mathrm{bat}}}}
\newcommand{\mkwdot}{\ensuremath{m_{\mathrm{kW},\cdot}}}

\newcommand{\etaPMAD}{\ensuremath{\eta_{\mathrm{PMAD}}}} 
\newcommand{\rhoDoD}{\ensuremath{\rho_{\mathrm{DoD}}}}   
\newcommand{\fPVEOL}{\ensuremath{f_{\mathrm{EOL,PV}}}}   
\newcommand{\fBatEOL}{\ensuremath{f_{\mathrm{EOL,bat}}}} 
\newcommand{\Torbit}{\ensuremath{T_{\mathrm{orb}}}}      
\newcommand{\Rae}{\ensuremath{R_{\oplus}}}                
\newcommand{\muE}{\ensuremath{\mu_{\oplus}}}             
\newcommand{\betaang}{\ensuremath{\beta}}                

\newcommand{\CRFdef}{\ensuremath{\CRF=\frac{r(1+r)^{n}}{(1+r)^{n}-1}}}

\newcommand{\yr}{\ensuremath{\mathrm{yr}}}

\newcommand{\LEO}{\textsc{LEO}}

\newcommand{\sigSB}{\ensuremath{\sigma_{\mathrm{SB}}}}

\newcommand{\Pit}{\ensuremath{P_{\mathrm{IT}}}}
\newcommand{\Ptot}{\ensuremath{P_{\mathrm{tot}}}}
\newcommand{\alphaOH}{\ensuremath{\alpha_{\mathrm{OH}}}}
\newcommand{\etaPV}{\ensuremath{\eta_{\mathrm{PV}}}}
\newcommand{\etaGeom}{\ensuremath{\eta_{\mathrm{geom}}}}
\newcommand{\etaSun}{\ensuremath{f_{\odot}}}
\newcommand{\Ssol}{\ensuremath{S_{0}}}
\newcommand{\epsrad}{\ensuremath{\epsilon_{\mathrm{rad}}}}
\newcommand{\etaview}{\ensuremath{\eta_{\mathrm{view}}}}
\newcommand{\Trad}{\ensuremath{T_{\mathrm{rad}}}}
\newcommand{\Arad}{\ensuremath{A_{\mathrm{rad}}}}
\newcommand{\Apv}{\ensuremath{A_{\mathrm{PV}}}}
\newcommand{\spPV}{\ensuremath{\mathrm{SP}_{\mathrm{PV}}}}
\newcommand{\sigRad}{\ensuremath{\sigma_{\mathrm{rad}}}}
\newcommand{\Lkg}{\ensuremath{L_{\$/\mathrm{kg}}}}
\newcommand{\Bkg}{\ensuremath{B_{\$/\mathrm{kg}}}}
\newcommand{\CRF}{\ensuremath{\mathrm{CRF}}}

\newcommand{\mkw}{\ensuremath{m_{\mathrm{kW}}}}

\usetikzlibrary{arrows.meta,positioning,fit}

\begin{document}

\title{Orbital Data Centers: Spacecraft Constraints and Economic Viability}

\author{Slava G. Turyshev}
\affiliation{
Jet Propulsion Laboratory, California Institute of Technology,\\
4800 Oak Grove Drive, Pasadena, CA 91109-0899, USA
}%

\date{\today}

\begin{abstract}
Orbital data centers are being evaluated as solar-powered compute constellations and relay-integrated processing platforms. Their feasibility is not set by orbital solar flux alone, but by simultaneous closure of photovoltaic generation, eclipse recharge, radiative heat rejection, sustained space-to-ground communications, utilization, replacement cadence, and delivered compute-years over finite mission life. This paper derives necessary cluster-level competitiveness conditions using delivered information-technology (IT) electrical power $P_{\rm IT}$, deployed mass per delivered IT power $m_{\rm kW}$ in kg/kW, communication intensity $\Gamma=D_{\rm sg}/E_{\rm IT}$, sustained communication ceiling $\Gamma_{\max}$, effective utilization $U_{\rm eff}$, and lifetime penalty $\Pi_{\rm life}$. For a representative $P_{\rm IT}=\SI{1}{MW}$ high-sunlight anchor, the base case gives beginning-of-life photovoltaic area $A^{\rm BOL}_{\rm PV}=\SI{5.64e3}{m^2}$, radiator area $A_{\rm rad}=\SI{2.50e3}{m^2}$, and $\SI{29.4}{kg/kW}$ for photovoltaic, storage, and radiator mass; fixed spacecraft mass raises the total to $34$--$59~\mathrm{kg/kW}$. At \(m_{\rm kW}\simeq 40~{\rm kg/kW}\), a terrestrial infrastructure benchmark of \(10\)--\(40~{\rm k\$/kW}\) allows only \(250\)--\(1000~{\rm \$/kg}\) for the combined launch and spacecraft-build cost before space-to-ground communications, operations, utilization, and lifetime terms are included. That allowance is $3.4$--$13.5$ times below the current public Falcon 9 dedicated low-Earth-orbit launch-price benchmark alone, before spacecraft build is included. Space-native preprocessing and communications-integrated edge compute are credible early regimes; terrestrial-user general compute closes only for low Earth-coupled communication intensity, high effective utilization, long delivered lifetime, and very low combined launch-plus-build cost.
\end{abstract}

\maketitle

\section*{Nomenclature}

Tables~\ref{tab:abbrev_terms} and~\ref{tab:core_symbols} summarize the non-common abbreviations, technical terms, and core symbols used in the paper. Symbols are also defined when they first enter the governing equations.

\begin{table*}[h!]
\caption{Abbreviations and technical terminology used in the paper.}
\label{tab:abbrev_terms}
\centering
\small
\setlength{\tabcolsep}{6pt}
\renewcommand{\arraystretch}{1.10}
\begin{tabular}{@{}l l@{}}
\toprule
Term & Definition used in this paper \\
\midrule
ADCS & Attitude determination and control system \\
AE9/AP9 & Trapped electron and proton radiation environment model suite \\
AI & Artificial intelligence; used only as a workload category \\
BOL / EOL & Beginning of life / end of life \\
CRF & Capital recovery factor \\
GEO / LEO & Geostationary Earth orbit / low Earth orbit \\
GNC & Guidance, navigation, and control \\
ISL & Inter-satellite link used for internal cluster traffic \\
IT power & Information-technology electrical power at the compute-payload interface \\
Node & Independently deployable spacecraft unit delivering specified EOL IT power \\
ODC & Orbital data center \\
PMAD & Power management and distribution \\
PV & Photovoltaic solar-electric generation subsystem \\
RF & Radio frequency \\
Space-to-ground link & Direct or relay-mediated communication path between orbit and ground \\
TID / SEE & Total ionizing dose / single-event effect \\
C1 & Space-native processing that reduces space-to-ground data or enables local action \\
C2 & Communications-integrated edge compute sharing relay or broadband infrastructure \\
C3 & Terrestrial-user general compute competing with terrestrial data centers \\
\bottomrule
\end{tabular}
\end{table*}

\begin{table*}[h!]
\caption{Core symbols used in the paper. Symbols are grouped by function for quick reference.}
\label{tab:core_symbols}
\centering
\small
\setlength{\tabcolsep}{4pt}
\renewcommand{\arraystretch}{1.08}
\begin{tabular}{@{}l l l l@{}}
\toprule
Symbol & Units & Meaning & Defined in \\
\midrule
\multicolumn{4}{@{}l}{\emph{Terrestrial demand and planning}}\\
$E_{\rm yr}$ & TWh yr$^{-1}$ & Annual electricity demand & Eq.~(\ref{eq:PavgPname}) \\
$P_{\rm avg}$ & GW & Average electrical load & Eq.~(\ref{eq:PavgPname}) \\
$P_{\rm name}$ & GW & Nameplate-equivalent provisioned power & Eq.~(\ref{eq:PavgPname}) \\
$U$ & -- & Utilization of installed IT or nameplate power & Eq.~(\ref{eq:PavgPname}) \\
$\chi$ & -- & Regional deliverability factor & Eq.~(\ref{eq:capadd_required}) \\
$CF_i$, $CC_i$ & -- & Capacity factor and peak capacity credit of resource $i$ & Eq.~(\ref{eq:capadd_required}) \\
$LF$ & -- & Load factor, $P_{\rm avg}/P_{\rm pk}$ & Eq.~(\ref{eq:capadd_required}) \\
\multicolumn{4}{@{}l}{\emph{Orbital node and subsystem sizing}}\\
$P_{\rm IT}$ & W & Delivered IT electrical power & Sec.~\ref{sec:boundary} \\
$P_{\rm tot}$ & W & Total spacecraft electrical power & Eq.~(\ref{eq:Ptot_def}) \\
$\alpha_{\rm OH}$ & -- & Orbital overhead factor, $P_{\rm tot}/P_{\rm IT}$ & Eq.~(\ref{eq:Ptot_def}) \\
$M_{\rm sys}$ & kg & Total launched and operated node mass & Eq.~(\ref{eq:mkw_def}) \\
$m_{\rm kW}$ & kg kW$^{-1}$ & Deployed mass per delivered IT kilowatt & Eq.~(\ref{eq:mkw_def}) \\
$m_{\rm kW}^{(\Sigma)}$ & kg kW$^{-1}$ & Cluster-averaged deployed mass per IT kilowatt & Eq.~(\ref{eq:mkw_cluster}) \\
$A_{\rm PV}^{\rm BOL}$ & m$^2$ & Beginning-of-life photovoltaic area & Eq.~(\ref{eq:Apv}) \\
$A_{\rm rad}$ & m$^2$ & Effective radiator area & Eq.~(\ref{eq:Arad}) \\
$SP_{\rm PV}$ & W kg$^{-1}$ & Deployed photovoltaic system specific power & Table~\ref{tab:key_params} \\
$t_{\rm ecl}$ & h & Eclipse duration per orbit & Sec.~\ref{sec:eclipse_storage} \\
$e_b$ & Wh kg$^{-1}$ & Battery specific energy & Table~\ref{tab:key_params} \\
$\rho_{\rm DoD}$ & -- & Allowable battery depth-of-discharge fraction & Sec.~\ref{sec:eclipse_storage} \\
$f_{\rm EOL,bat}$ & -- & End-of-life usable battery capacity fraction & Sec.~\ref{sec:eclipse_storage} \\
$f_{\rm EOL,PV}$ & -- & End-of-life photovoltaic power fraction & Sec.~\ref{sec:pv_sizing} \\
$T_{\rm rad}$ & K & Radiator operating temperature & Sec.~\ref{sec:thermal} \\
$\sigma_{\rm rad}$ & kg m$^{-2}$ & Effective radiator areal density & Table~\ref{tab:key_params} \\
\multicolumn{4}{@{}l}{\emph{Communications and workload coupling}}\\
$D_{\rm sg}$ & bit & Bidirectional data crossing the space-to-ground interface & Eq.~(\ref{eq:Gamma}) \\
$E_{\rm IT}$ & J & IT energy consumed by workload & Eq.~(\ref{eq:Gamma}) \\
$R_{\rm sg}$ & bit s$^{-1}$ & Sustained average space-to-ground rate & Eq.~(\ref{eq:Ravg}) \\
$\Gamma$ & bit J$^{-1}$ & Communication intensity, $D_{\rm sg}/E_{\rm IT}$ & Eq.~(\ref{eq:Gamma}) \\
$\Gamma_{\rm ext}$ & bit J$^{-1}$ & Space-to-ground portion of workload intensity & Eq.~(\ref{eq:Gamma_ext_fraction}) \\
$\Gamma_{\max}$ & bit J$^{-1}$ & Sustained communication-intensity ceiling & Eq.~(\ref{eq:Gamma_max}) \\
\multicolumn{4}{@{}l}{\emph{Lifecycle and economics}}\\
$U_{\rm eff}$ & -- & Effective utilization after availability penalties & Eq.~(\ref{eq:Util_eff}) \\
$\lambda_{\rm eff}$ & yr$^{-1}$ & Effective service-loss hazard rate & Eq.~(\ref{eq:lambda_eff}) \\
$\Pi_{\rm life}$ & -- & Delivered-compute lifetime penalty & Eq.~(\ref{eq:Pi_life}) \\
$L_{\$/\mathrm{kg}}$ & \$ kg$^{-1}$ & Launch and delivery cost to operational orbit & Sec.~\ref{sec:orbit_capitalcost} \\
$B_{\$/\mathrm{kg}}$ & \$ kg$^{-1}$ & Spacecraft build, integration, and qualification cost & Sec.~\ref{sec:orbit_capitalcost} \\
$C_{\rm terr}$ & \$ kW$^{-1}$ & Terrestrial infrastructure benchmark per IT kilowatt & Eq.~(\ref{eq:Cground}) \\
$C_{\rm link}$ & \$ kW$^{-1}$ & Space-to-ground communications and terminal cost & Eq.~(\ref{eq:corbit}) \\
$C_{\rm ops}$ & \$ kW$^{-1}$ & Capital-cost-equivalent operations and lifecycle cost & Eq.~(\ref{eq:corbit}) \\
\bottomrule
\end{tabular}
\end{table*}




\section{Introduction}
\label{sec:intro}

Large computing facilities are increasingly constrained by the rate at which deliverable electrical infrastructure can be added. Recent U.S. assessments estimate data-center electricity use at \(176~\mathrm{TWh\,yr^{-1}}\) in 2023 and project \(325\text{--}580~\mathrm{TWh\,yr^{-1}}\) by 2028, corresponding to average load rising from \(20.1~\mathrm{GW}\) to \(37.1\text{--}66.2~\mathrm{GW}\) \cite{LBNL2024DataCenters}. Reliability assessments identify data centers and other large loads as material adequacy and deliverability drivers for the bulk power system \cite{NERC2024LTRA}. The global trend is consistent: data-center electricity growth is concentrated in load pockets, so the binding constraint is often interconnection, transmission, firm capacity, and equipment lead time rather than gross annual energy availability \cite{IEA2025EnergyAI}. The terrestrial calibration used below is therefore a U.S.-context time-to-power benchmark. The spacecraft sizing equations are geography-independent; the economic comparison is not. Applying the model outside this U.S. benchmark requires local values of the effective deliverability factor \(\chi\), the terrestrial infrastructure benchmark \(\Cterr\), electricity price, permitting delay, and service-level requirements.

Orbital data centers (ODCs) have moved from conceptual energy arguments to explicit spacecraft-system proposals. Public studies and filings now describe solar-powered compute constellations, high-sunlight orbit selection, optical inter-satellite links (ISLs), gateway specialization, commercial relay-service integration, and modular node counts as design variables \cite{GoogleSuncatcherPaper2025,Bargatin2025TetherODC,ESPISBDC2025,TUMSBDC2025,NASACommServices2026,StarlinkProgress2025,FCCSpaceXODC2026,NVIDIASpaceComputing2026}. The technical question is whether the substituted orbital constraint set is more favorable than the terrestrial one for a specified workload and service level. Continuous compute in orbit is a specific-power and eclipse-storage problem because photovoltaic (PV) arrays must supply the regulated bus and recharge storage. Thermal closure is a radiative-area problem because vacuum provides no convective heat sink. Terrestrial-user computation is a sustained-throughput problem because visibility, weather, gateway scheduling, relay allocation, and pointing losses reduce peak optical or radio-frequency link rates to lower average service rates. Economic performance is a delivered-compute-years problem because radiation, debris, drag, replacement delay, and processor obsolescence determine how much installed compute is delivered over mission life.

This paper develops a physics-based competitiveness model for that substituted constraint set. The contribution is a cluster-level viability condition that maps delivered information-technology (IT) power into three coupled quantities: deployed mass per delivered kilowatt (kW), admissible space-to-ground data per unit compute energy, and delivered compute-years after utilization and lifetime penalties. The model separates the subsystem floor--PV generation, eclipse storage, radiator burden, and overhead power--from architecture-sensitive choices such as node granularity, gateway specialization, reserve policy, space-to-ground communications link realization, and fault-domain design. This separation is essential because current ODC architectures differ primarily in architecture, while all must satisfy the same power, thermal, communication, and lifetime constraints.

For the constellation-first architecture adopted below, terrestrial-user general compute must satisfy
\begin{equation}
(\Lkg+\Bkg)\mkw^{(\Sigma)}+\CGS^{(\Sigma)}+C_{\rm ops}^{(\Sigma)} \lesssim \Cterr,
\qquad
\Gamma \le \Gamma_{\max}^{(\Sigma)},
\label{eq:C3_joint}
\end{equation}
where \(\mkw^{(\Sigma)}\) is cluster-averaged deployed mass per delivered IT kilowatt, \(\Gamma\equiv D_{\rm sg}/E_{\rm IT}\) is the space-to-ground communication intensity of the workload, \(\Gamma_{\max}^{(\Sigma)}\) is the sustained space-to-ground communications ceiling after geometry, weather, scheduling, and gateway allocation, \(\CGS^{(\Sigma)}\) is the cluster-allocated space-to-ground communications and terminal cost, and \(C_{\rm ops}^{(\Sigma)}\) is the capital-cost-equivalent lifecycle term not proportional to launched mass. Eq.~(\ref{eq:C3_joint}) is the decision rule used throughout the paper. A terrestrial-user orbital cluster must close mass per delivered kW, space-to-ground data per unit compute, and delivered compute-years simultaneously; closure of only one term is not sufficient.

The current ODC discussion separates into three technically distinct workload computing regimes, denoted C1--C3 in this paper. C1 is space-native processing: compute is placed near an orbital sensor, spacecraft, or control loop to reduce downlink volume or enable local action. C2 is communications-integrated edge compute: processing is hosted inside a relay or broadband constellation, where pointing, gateway access, crosslinks, and operations are already mission-critical. C3 is terrestrial-user general compute: the orbital system competes with terrestrial data centers for workloads whose users and storage are primarily on Earth. Google's \emph{Suncatcher} study, the European Space Policy Institute/Technical University of Munich (ESPI/TUM) gigawatt-class cost analysis, the SpaceX filing with the U.S. Federal Communications Commission (FCC), the NASA commercial relay-services program, Starlink's published optical inter-satellite-link (ISL) architecture, NVIDIA space-computing hardware, tethered ODC architecture studies, and recent in-orbit learning and compute-placement work all occupy this design space \cite{GoogleSuncatcherPaper2025,Bargatin2025TetherODC,ESPISBDC2025,TUMSBDC2025,NASACommServices2026,StarlinkProgress2025,FCCSpaceXODC2026,NVIDIASpaceComputing2026,Chabra2026OrbitalBrain,ThummalaFalco2025ComputeInSpace}. The Starlink reference is used here as evidence of commercial low-Earth-orbit optical crosslink infrastructure and NASA-demonstrated relay-service use, not as an assumption that Starlink provides direct optical space-to-ground service for ODCs. Earlier work framed space-based data centers as a data-processing architecture \cite{PeriolaKolawole2019SBDC}; the present paper supplies the coupled mass-throughput-lifetime-economic closure needed to assess which operating regimes can close.

The paper is organized as follows. Section~\ref{sec:power_need} converts terrestrial growth projections into an energy-, adequacy-, and deliverability-limited power-delivery benchmark. Section~\ref{sec:system} defines the node boundary, distributed reference architecture, fleet-scale translation, workload regimes, and core metrics. Sections~\ref{sec:power}--\ref{sec:mkw} derive the subsystem floor through solar generation, eclipse storage, thermal rejection, and the mass-per-power decomposition. Section~\ref{sec:comms} develops the sustained space-to-ground communications ceiling and traffic asymmetry. Sections~\ref{sec:radiation}--\ref{sec:econ} propagate radiation, lifetime, replenishment, and operations into delivered compute-years and break-even boundaries, including dedicated-site and relay-service space-to-ground link realizations. Section~\ref{sec:business_cases} interprets the results across the three workload regimes, Section~\ref{sec:limitations} defines the validation path and model applicability, and Section~\ref{sec:conclusion} states the conclusions.

\section{Terrestrial benchmark: time to deliver power}
\label{sec:power_need}

The terrestrial comparison is expressed in the variables that determine deployment pace: annual energy, credited capacity during system-stress periods, and deliverability to the load pocket on the commercial schedule. Here \emph{time to deliver power} means the interval required to add firm, interconnected, geographically deliverable electrical capacity at the bus or zone where the computing load is sited. This section converts projected data-center demand growth into that form so that the terrestrial benchmark is commensurate with orbital infrastructure.

An incremental annual energy requirement \(\Delta E\) corresponds to an incremental average load
\begin{equation}
\Delta P_{\rm avg}=\frac{\Delta E}{8760~\mathrm{h}}
\approx 0.114~\mathrm{GW}\left(\frac{\Delta E}{1~\mathrm{TWh\,yr^{-1}}}\right),
\label{eq:Delta-Pavg}
\end{equation}
which allows direct comparison to generation build rates. Using the LBNL projection range, U.S.\ data-center electricity use increases from \(E_{2023}\simeq176~\mathrm{TWh}\) to \(E_{2028}\simeq325\text{--}580~\mathrm{TWh}\), implying
\begin{equation}
\Delta P_{\rm avg}\simeq 17.0\text{--}46.1~\mathrm{GW}
\quad\text{over the 2023--2028 window.}
\label{eq:dPavg_range}
\end{equation}
Over a five-year planning horizon this corresponds to a delivered average-load ramp of
\begin{equation}
\dot P_{\rm avg}\equiv \frac{\Delta P_{\rm avg}}{\Delta t}
\simeq (3.4\text{--}9.2)~\mathrm{GW\,yr^{-1}}
\left(\frac{5~\yr}{\Delta t}\right),
\label{eq:dPavg_rate}
\end{equation}
which is the relevant \emph{time-to-power} benchmark \cite{NERC2024LTRA}.

For later comparison to orbital nodes, the corresponding average and nameplate-equivalent electrical powers are
\begin{equation}
\Pavg \equiv \frac{\Eyr}{8760~\mathrm{h}},
\qquad
\Pname \equiv \frac{\Eyr}{8760~\mathrm{h}\,\Util}=\frac{\Pavg}{\Util},
\label{eq:PavgPname}
\end{equation}
so that at \(\Util=0.5\) the 2028 high case corresponds to \(\Pname\simeq132.4~\mathrm{GW}\) of nameplate-equivalent provisioned capacity.

A compact supply-response expression for resource \(i\) is
\begin{equation}
\Delta P_{i}^{(\mathrm{req})}
=
\Delta P_{\rm avg}\,
\max\!\left[
\frac{1}{\chi\,\mathrm{CF}_i},\;
\frac{1}{\chi\,\mathrm{LF}\,\mathrm{CC}_i}
\right],
\label{eq:capadd_required}
\end{equation}
where \(\chi\in(0,1]\) is an effective deliverability factor, \(\mathrm{CF}_i\) the energy capacity factor, \(\mathrm{CC}_i\) the peak capacity credit, and \(\mathrm{LF}\equiv P_{\rm avg}/P_{\rm pk}\) the data-center load factor. Eq.~(\ref{eq:capadd_required}) makes explicit why the terrestrial comparator is not annual energy alone but the ability to add deliverable and credited capacity on the Eq.~(\ref{eq:dPavg_rate}) timeline. The orbital proposition attempts to bypass precisely this deliverability constraint, at the cost of substituting the mass, space-to-ground link, and lifetime constraints developed below.

\section{System definition and core metrics}
\label{sec:system}

\subsection{System boundary: node definition and accounting interfaces}
\label{sec:boundary}

We model an orbital data center (ODC) as a set of \emph{nodes}, where a node is the smallest independently deployable unit that delivers a specified end-of-life (EOL) information technology (IT) electrical power \Pit\ to a compute payload at utilization \Util\ over an operational lifetime \(T\). The node boundary includes \emph{all} subsystems that must be launched and operated in orbit to sustain \Pit: photovoltaics (PV), eclipse storage (if any), power conditioning and distribution, thermal transport and radiators, attitude determination and control, propulsion for station-keeping and collision avoidance, communications terminals, flight avionics, structure/deployment mechanisms, radiation shielding margin, and compute packaging. This boundary is intentionally strict: it ensures that the mass-per-power metric \mkw\ and the cost discriminator in Sec.~\ref{sec:econ} are defined by explicit accounting rather than by implicit exclusions.

We define \Pit\ at the electrical interface feeding the IT payload (accelerators, memory, and local switching),
and \Ptot\ as the total node electrical power that must be generated and conditioned. Unless stated otherwise, all sizing relations are interpreted as EOL constraints; beginning-of-life (BOL) oversizing to accommodate degradation is treated explicitly in Sec.~\ref{sec:power}.

\subsection{Reference architecture, operations, and lifecycle accounting}
\label{sec:arch_layers}

We decompose an orbital data center into three coupled layers: the \emph{physical architecture}, the
\emph{operational architecture}, and the \emph{lifecycle architecture}. The physical architecture comprises compute nodes,
gateway nodes, inter-satellite networking, and the space-to-ground communications segment. The operational architecture specifies task
placement, checkpoint/restart doctrine, gateway scheduling, autonomy, collision avoidance, and software update policy.
The lifecycle architecture covers manufacture and qualification, launch and deployment, commissioning, nominal
operations, replenishment, retirement, and disposal. This separation matters because the same node-level physics floor
can map into different delivered-compute economics depending on how the latter two layers are implemented.

This distinction also defines the maintenance and deployment assumptions used in the cost model. Hardware maintenance is represented by telemetry-based health management, software update policy, orbit maintenance, reserve allocation, and replacement cadence; crewed on-orbit servicing is not assumed. Deployment denotes manufacture, qualification, launch, commissioning, and disposal of modular nodes rather than assembly of a crewed orbital facility. This convention matches the cost structure used in Secs.~\ref{sec:radiation}--\ref{sec:econ}, where the dominant economic terms are mass-scaled hardware cost, space-to-ground communications cost, and lifecycle/replenishment cost.

\emph{Physical architecture.} Compute nodes, gateway nodes, PV, storage, radiators, communications terminals, and inter-satellite plus space-to-ground link elements determine the first-order mass and throughput quantities \(m_{\rm kW}\), \(m_{\rm kW,fixed}\), \(\Gamma_{\max}\), and \(\alphaOH\). These enter the mass-scaled capital-cost term through \((\Lkg+\Bkg)m_{\rm kW}\).

\emph{Operational architecture.} Task placement, checkpoint/restart doctrine, gateway scheduling, autonomy, conjunction response, and update policy determine \(U_{\rm eff}\), \(f_{\rm sched}\), and the sustained space-to-ground rate \(\overline{R}_{\rm sg}\). These map into delivered-compute penalty and the non-mass operations term \(C_{\rm ops}\).

\emph{Lifecycle architecture.} Manufacture and qualification, launch and deployment, commissioning, replenishment, retirement, and disposal determine \(\lambda_{\rm eff}\), \(\tau_{\rm rep}\), reserve fraction, and mission life. These map into \(\Pi_{\rm life}\), replacement delay, effective launch and spacecraft build burden, and disposal reserve.

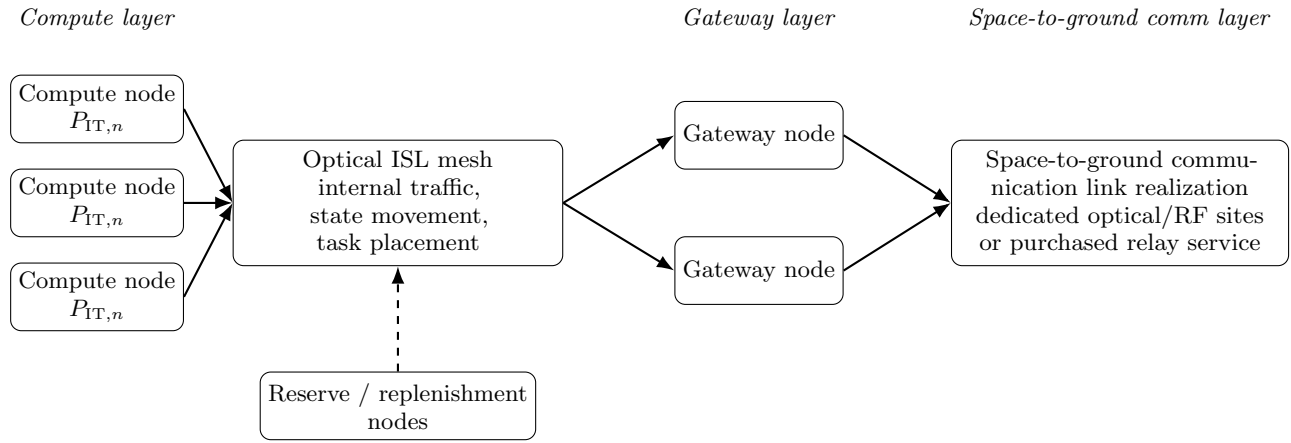
\begin{figure*}[t]
\centering
\begin{tikzpicture}[
smallbox/.style={draw, rounded corners, align=center, minimum width=2.2cm, minimum height=0.9cm, inner sep=3pt},
midbox/.style={draw, rounded corners, align=center, text width=4.1cm, minimum height=1.35cm, inner sep=4pt},
earthbox/.style={draw, rounded corners, align=center, text width=4.2cm, minimum height=1.35cm, inner sep=4pt},
line/.style={-Latex, thick},
dashedline/.style={-Latex, thick, dashed}
]
\node[smallbox] (c1) at (0.0,1.25) {Compute node\\ \Pitn};
\node[smallbox] (c2) at (0.0,0.0) {Compute node\\ \Pitn};
\node[smallbox] (c3) at (0.0,-1.25) {Compute node\\ \Pitn};

\node[midbox] (isl) at (4.0,0.0) {Optical ISL mesh\\internal traffic, state movement,\\task placement};

\node[smallbox] (g1) at (8.8,0.9) {Gateway node};
\node[smallbox] (g2) at (8.8,-0.9) {Gateway node};

\node[earthbox] (earth) at (13.6,0.0) {Space-to-ground communication link realization\\dedicated optical/RF sites\\or purchased relay service};

\node[smallbox] (reserve) at (4.0,-2.7) {Reserve / replenishment\\nodes};

\draw[line] (c1.east) -- (isl.west);
\draw[line] (c2.east) -- (isl.west);
\draw[line] (c3.east) -- (isl.west);
\draw[line] (isl.east) -- (g1.west);
\draw[line] (isl.east) -- (g2.west);
\draw[line] (g1.east) -- (earth.west);
\draw[line] (g2.east) -- (earth.west);
\draw[dashedline] (reserve.north) -- (isl.south);

\node[align=center] at (0.0,2.45) {\emph{Compute layer}};
\node[align=center] at (8.8,2.45) {\emph{Gateway layer}};
\node[align=center] at (13.6,2.45) {\emph{Space-to-ground comm layer}};
\end{tikzpicture}
\caption{Reference distributed ODC architecture used throughout the paper. Compute nodes are linked by optical inter-satellite links (ISL), a smaller set of gateway nodes concentrates space-to-ground traffic, and the space-to-ground communications term may be realized either as dedicated ground infrastructure or as purchased relay service. Reserve and replenishment are treated at cluster level through the lifecycle model rather than through an assumed on-orbit servicing doctrine.}
\label{fig:odc_reference_arch}
\end{figure*}

\subsection{Fleet-scale translation of the terrestrial benchmark}
\label{sec:fleet_scale}

The translation from node-level sizing to system-scale deployment should be made explicit rather than left implicit.
For a target aggregate IT power \(P_{\rm IT,\Sigma}^{\rm targ}\), the required active compute-node count is
\begin{equation}
N_c = \frac{P_{\rm IT,\Sigma}^{\rm targ}}{\Pitn}.
\label{eq:Nc_target}
\end{equation}
The scale implication follows directly from Eq.~(\ref{eq:Nc_target}). For example, \(P_{\rm IT,\Sigma}^{\rm targ}=\SI{1}{GW}\) requires 10,000 active \(100~\mathrm{kW}\) compute nodes or 1,000 active \(1~\mathrm{MW}\) compute nodes. Applying Eq.~(\ref{eq:Nc_target}) to the \(17.0\)--\(46.1~\mathrm{GW}\) incremental average-load benchmark from Eq.~(\ref{eq:dPavg_range}) gives 170,000--461,000 active \(100~\mathrm{kW}\) nodes or 17,000--46,100 active \(1~\mathrm{MW}\) nodes before reserve, gateway, and replacement margin. Table~\ref{tab:fleet_scale} gives the corresponding values for representative node powers; it is a lower bound on deployed spacecraft count because reserve and gateway spacecraft are excluded.

\begin{table*}[t]
\caption{Fleet-scale translation from aggregate orbital IT power to active compute-node count. The 17.0--46.1 GW range is the incremental average-load benchmark used in the terrestrial time-to-power calibration. Counts exclude reserve nodes, gateway nodes, and replacement margin.}
\label{tab:fleet_scale}
\centering
\small
\setlength{\tabcolsep}{6pt}
\renewcommand{\arraystretch}{1.15}
\begin{tabular}{@{}l r r r r@{}}
\toprule
Node power \(\Pitn\) & 1 GW & 17.0 GW & 46.1 GW & Node loss \\
\midrule
\SI{10}{kW}  & 100,000 & 1,700,000 & 4,610,000 & 0.001\% \\
\SI{100}{kW} & 10,000  & 170,000   & 461,000   & 0.01\% \\
\SI{1}{MW}   & 1,000   & 17,000    & 46,100    & 0.1\% \\
\SI{10}{MW}  & 100     & 1,700     & 4,610     & 1.0\% \\
\bottomrule
\end{tabular}
\end{table*}

Two conclusions follow immediately. First, the \(\Pit=\SI{1}{MW}\) case is best interpreted as a subsystem-physics
scaling anchor, not as a claim that an early deployment would favor \SI{1}{MW} monolithic nodes. The \(\Pitn=\SI{100}{kW}\)
and \(\Pit=\SI{1}{MW}\) cases used later are analytic power anchors, not claims of current flight-qualified node
classes; they are retained to expose scaling in \(\mkw^{(\Sigma)}\), \(\Gamma_{\max}^{(\Sigma)}\), and replacement granularity across duplication-dominated and communications-limited regimes. Second, node
granularity is inseparable from service continuity: reducing \(\Pitn\) increases duplication and fleet size, but it also
reduces the capacity fraction removed by a single node loss. The selected node power therefore sits inside a trade
among duplication, gateway scaling, reserve policy, and allowable correlated-loss fraction, not inside a single scalar optimum.

\subsection{Workload regimes and technical definitions}
\label{sec:classes}

The labels C1, C2, and C3 distinguish how orbital location and infrastructure sharing modify the terms in Eq.~(\ref{eq:C3_joint}). They are not independent design models and they are not assumed to be a standard taxonomy. Each class is evaluated with the same mass, space-to-ground link, utilization, and lifetime constraints.

C1 denotes \emph{space-native processing}: the source data or control loop is already in orbit, and computation is valuable because it reduces the data volume delivered to Earth, increases the information content per downlinked bit, or enables local action. In the model, C1 primarily reduces space-to-ground communication traffic \(D_{\rm sg}\) and therefore communications intensity \(\Gamma_{\rm ext}\). C2 denotes \emph{communications-integrated edge compute}: the compute function is hosted as an incremental payload or software function inside a relay or broadband constellation, so part of the bus, pointing, gateway, crosslink, and operations burden is allocated to the primary communications service rather than to compute alone. C3 denotes \emph{terrestrial-user general compute}: users, storage, and value capture are primarily terrestrial, so the cluster must satisfy the full condition in Eq.~(\ref{eq:C3_joint}) without a large location-value credit.

This classification is technically useful because the dominant feasibility mechanism differs across the three cases. C1 changes the workload term; C2 changes the allocation of fixed mass and space-to-ground communications cost; C3 changes neither and is therefore the stress case. Onboard autonomy and remote-sensing preprocessing are C1-like, compute embedded in relay or broadband networks is C2-like, and solar-powered batch-compute constellations serving terrestrial users are C3-like unless most data are generated and consumed in orbit. Table~\ref{tab:classes} summarizes the engineering discriminators.

\begin{table*}[t]
\caption{Operating regimes and dominant feasibility discriminators. The C1--C3 labels are author-defined operating regimes evaluated by the same mass, communications, utilization, and lifetime model. C1 reduces the Earth-coupled data term, C2 reallocates shared spacecraft and communications burden, and C3 must satisfy the full joint condition in Eq.~(\ref{eq:C3_joint}).}
\label{tab:classes}
\centering
\small
\setlength{\tabcolsep}{4pt}
\renewcommand{\arraystretch}{1.12}
\begin{tabular}{@{}lll@{}}
\toprule
Regime & Technical definition & Governing discriminator \\
\midrule
C1 & \parbox[t]{0.44\textwidth}{\raggedright Space-native processing: the data source or control loop is already in orbit; computation transforms raw data into lower-volume or higher-value products, or enables local action before Earth contact.} & \parbox[t]{0.43\textwidth}{\raggedright Reduction in space-to-ground traffic, represented by lower $D_{\rm sg}$ and lower $\Gamma_{\rm ext}$, compared with downlinking raw data or waiting for ground-based decisions.} \\
\noalign{\vskip 2pt}
C2 & \parbox[t]{0.44\textwidth}{\raggedright Communications-integrated edge compute: compute is an incremental payload or software function on a relay or broadband architecture with existing pointing, gateway, crosslink, and operations infrastructure.} & \parbox[t]{0.43\textwidth}{\raggedright Allocated fractions of $\mkwfixed$, $\CGS$, and $C_{\rm ops}$, plus the condition that the incremental workload does not drive additional space-to-ground link duty factor or gateway count.} \\
\noalign{\vskip 2pt}
C3 & \parbox[t]{0.44\textwidth}{\raggedright Terrestrial-user general compute: users and data stores are primarily terrestrial; orbital compute competes directly with terrestrial data-center infrastructure.} & \parbox[t]{0.43\textwidth}{\raggedright Simultaneous satisfaction of low $\mkw$, $\Gamma\le\Gamma_{\max}$, high $U_{\rm eff}$, and acceptable delivered compute-years.} \\
\bottomrule
\end{tabular}
\end{table*}

\subsection{A workload-to-architecture coupling metric: communication intensity}
\label{sec:Gamma}

For Earth-serving workloads (C3), orbit competes with terrestrial compute while satisfying an additional space--ground data-transfer constraint. The corresponding workload metric is the \emph{communication intensity}, defined as space-to-ground data exchanged per unit IT energy:
\begin{equation}
\Gamma \equiv \frac{D_{\mathrm{sg}}}{E_{\mathrm{IT}}}\quad [\mathrm{bit/J}],
\label{eq:Gamma}
\end{equation}
where \(D_{\mathrm{sg}}\) is the net bidirectional space-to-ground traffic attributable to the workload
(payload I/O, checkpoints, model/weights transfer as applicable), and \(E_{\mathrm{IT}}\) is IT energy consumed.

For a node delivering \Pit\ at utilization \Util\ and sustaining an average bidirectional space-to-ground rate
\(\overline{R}_{\mathrm{sg}}\), the implied communication intensity is
\begin{equation}
\Gamma \simeq \frac{\overline{R}_{\mathrm{sg}}}{\Util\,\Pit}.
\label{eq:Gamma2}
\end{equation}
Eq.~(\ref{eq:Gamma2}) makes communications a verifiable feasibility constraint: for fixed \(\Pit\) and \Util,
a required \(\overline{R}_{\mathrm{sg}}\) implies a minimum \(\Gamma\) that must be supportable under duty-cycle and weather limits
(Sec.~\ref{sec:comms_sanity}).

For workloads with in-orbit reuse, caching, or staged datasets, only a fraction \(f_{\rm ext}\in[0,1]\) of the gross
workload traffic need cross the space-to-ground link. If \(\Gamma_{\rm gross}\) denotes the total workload communication
intensity and \(\Gamma_{\rm ext}\) the space-to-ground portion, then
\begin{equation}
\Gamma_{\rm ext} = f_{\rm ext}\,\Gamma_{\rm gross}.
\label{eq:Gamma_ext_fraction}
\end{equation}
The communications ceiling derived later should therefore be applied to \(\Gamma_{\rm ext}\), not to purely internal
cluster traffic that remains on inter-satellite links. This distinction matters for pre-positioned datasets, cached
model weights, and batch workloads whose dominant data motion is internal to the orbital cluster.

For engineering interpretation, convert \(\Gamma\) into data-per-energy units commonly used in systems practice.
Since \(1~\mathrm{kWh}=3.6\times 10^{6}~\mathrm{J}\),
\begin{equation}
\left(\frac{D_{\mathrm{sg}}}{E_{\mathrm{IT}}}\right)_{\mathrm{GB/kWh}}
\simeq 4.5\times 10^{-4}\,\Gamma.
\label{eq:Gamma_GBperKWh}
\end{equation}
As a concrete scaling, for a \(\Pit=1~\mathrm{MW}\) node at \(\Util=0.5\),
\begin{equation}
\Gamma \approx 2\times 10^{4}\ \mathrm{bit/J}\,
\left(\frac{\overline{R}_{\mathrm{sg}}}{10~\mathrm{Gb/s}}\right)
\left(\frac{0.5}{\Util}\right)
\left(\frac{1~\mathrm{MW}}{\Pit}\right),
\label{eq:Gamma_scale}
\end{equation}
corresponding to \(\sim 9~\mathrm{GB/kWh}\) of bidirectional space-to-ground traffic. Large-\(\Gamma\) workloads (interactive, storage-coupled services) become link- and availability-limited in orbit, whereas small-\(\Gamma\) workloads (compute-dense, delay-tolerant jobs; long training runs with pre-positioned data; space-native preprocessing) are structurally favored because they fit beneath the duty-cycle throughput ceiling derived in Sec.~\ref{sec:comms_sanity}.

\subsection{Power accounting and the mass-per-power metric}
\label{sec:mkW_def}

We define \Pit\ as delivered IT electrical power (accelerators, memory, local switching) at the payload power interface. The node must supply a higher total electrical power
\begin{equation}
\Ptot = \alphaOH \Pit,
\label{eq:Ptot_def}
\end{equation}
where \(\alphaOH \ge 1\) accounts for all non-IT loads required to sustain operation (power conditioning and distribution, thermal transport, pointing/jitter control, propulsion housekeeping, and communications terminals as applicable). 

The overhead factor \(\alphaOH\) plays the role that terrestrial power usage effectiveness (PUE) plays on the ground, but unlike terrestrial PUE it is explicitly orbit- and architecture-dependent: raising radiator temperature, tightening pointing, or increasing collision-avoidance activity can increase \(\alphaOH\) and thereby increase both PV and radiator sizing.

A compact feasibility scalar is the EOL deployed mass per delivered IT power,
\begin{equation}
m_{\mathrm{kW}} \equiv \frac{M_{\mathrm{sys}}}{P_{\mathrm{IT}}/(1~\mathrm{kW})}
= 1000\,\frac{M_{\mathrm{sys}}}{P_{\mathrm{IT}}}
\qquad [\mathrm{kg\,kW^{-1}}].
\label{eq:mkw_def}
\end{equation}
where \(M_{\mathrm{sys}}\) includes all launched and operated mass within the node boundary in
Sec.~\ref{sec:boundary}. This definition is chosen because the paper's first-order orbital infrastructure benchmark scales as \((\Lkg+\Bkg)\mkw\) (Sec.~\ref{sec:econ}), while the dominant physics-driven contributors to
\(M_{\rm sys}\) (PV, storage, radiators, and fixed spacecraft mass) scale approximately linearly
with \(P_{\rm tot}\) and therefore with \(P_{\rm IT}\).
The paper makes each contributor to \mkw\ explicit (Secs.~\ref{sec:power}--\ref{sec:mkw}). These later sections retain this benchmark form as the primary discriminator and use the effective-cost form only as a refinement for delivery overhead, reserve fraction, and industrial cost allocation.

The dependency structure used throughout the paper is straightforward: delivered IT power sets total platform power, subsystem constraints determine \(\mkw\), \(\Gamma_{\max}\), and \(\Pi_{\rm life}\), and these quantities then define the competitiveness boundary.

\begin{figure*}[t]
\centering
\begin{tikzpicture}[
  font=\small,
  box/.style={draw, rounded corners, align=center, inner sep=4pt, minimum height=11mm},
  widebox/.style={draw, rounded corners, align=left, inner sep=5pt},
  arr/.style={-{Latex[length=2mm]}, thick},
  node distance=5mm
]
\node[box,text width=20mm] (pit) {\(\Pit\)\\delivered\\IT power};
\node[box,text width=25mm,right=of pit] (ptot) {\(\Ptot=\alphaOH\Pit\)\\total platform\\power};
\node[widebox,text width=31mm,right=of ptot] (subsys) {subsystem constraints\\[2pt]
\(\bullet\) PV sizing\\
\(\bullet\) battery sizing\\
\(\bullet\) radiator sizing\\
\(\bullet\) comms ceiling\\
\(\bullet\) lifetime penalty};
\node[widebox,text width=28mm,right=of subsys] (metrics) {governing metrics\\[2pt]
\(\bullet\ \mkw\)\\
\(\bullet\ \Gamma_{\max}\)\\
\(\bullet\ \Pi_{\rm life}\)};
\node[widebox,text width=36mm,right=of metrics] (econ) {competitiveness boundary\\[2pt]
\(\Corbit \simeq (\Lkg+\Bkg)\mkw+\CGS+C_{\rm ops}\)\\
C3 requires \(\Gamma\le\Gamma_{\max}\)\\
delivered-cost penalty \(\propto \Pi_{\rm life}/U_{\rm eff}\)};
\draw[arr] (pit) -- (ptot);
\draw[arr] (ptot) -- (subsys);
\draw[arr] (subsys) -- (metrics);
\draw[arr] (metrics) -- (econ);
\end{tikzpicture}
\caption{Schematic overview of the competitiveness model. Delivered IT power sets total platform power through \(\alphaOH\). Subsystem constraints determine the three governing quantities that matter economically--\(\mkw\), \(\Gamma_{\max}\), and \(\Pi_{\rm life}\)--which then define the orbital competitiveness boundary.}
\label{fig:logic_map}
\end{figure*}
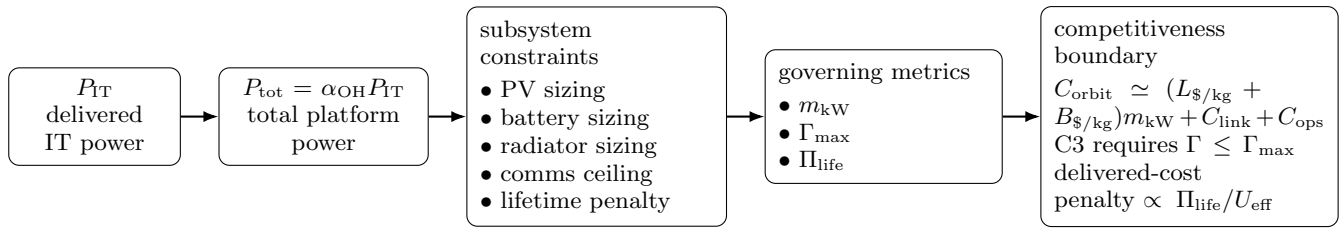

\section{Power subsystem sizing}
\label{sec:power}

\subsection{PV area and mass: continuous-load sizing couples to eclipse recharge}
\label{sec:pv_sizing}

We distinguish (i) the \emph{bus} power required by the platform, $\Ptot=\alphaOH\Pit$, Eq.~(\ref{eq:Ptot_def}), from (ii) the \emph{sunlit} PV electrical output that must simultaneously run the load and recharge storage. Let $\etaPMAD$ denote the end-to-end efficiency from PV output through power conditioning, regulation, and distribution to the regulated bus (including MPPT and conversion losses). Table~\ref{tab:key_params} records the representative technology envelope used throughout the sizing discussion. The purpose of the table is not to define a unique design point, but to anchor the physics in current and projected technology ranges for deployed solar arrays, storage, thermal hardware, and optical downlink capability.

\begin{table*}[t]
\caption{Subsystem parameter ranges and technology anchors used in the numerical examples. The values are architecture-level sizing inputs; MW-class ODC hardware requires mission-specific qualification and environmental analysis.}
\label{tab:key_params}
\centering
\small
\setlength{\tabcolsep}{5pt}
\renewcommand{\arraystretch}{1.12}
\begin{tabular}{@{}l c l@{}}
\toprule
Quantity & Symbol & Representative range or anchor \\
\midrule
Solar irradiance at 1 AU & \(\Ssol\) & \(\SI{1361}{W\,m^{-2}}\) \cite{KoppLean2011TSI} \\
Sunlight duty fraction & \(\etaSun\) & 0.6--0.98, from orbit geometry and beta angle \\
Deployed PV system specific power & \(\spPV\) & 30--165 W/kg spacecraft-survey anchor \cite{NASASOA2024Power} \\
PV-to-bus electrical efficiency & \(\etaPMAD\) & 0.88--0.95, including MPPT, conversion, and distribution \\
Battery specific energy & \(e_b\) & 150--265 Wh/kg spacecraft-survey anchor \cite{NASASOA2024Power} \\
Battery round-trip efficiency & \(\eta_b\) & 0.85--0.95, chemistry and temperature dependent \\
Usable depth of discharge & \(\rhoDoD\) & 0.3--0.8, set by cycle-life requirement \\
PV EOL/BOL power ratio & \(\fPVEOL\) & 0.75--0.95, orbit/radiation/lifetime dependent \\
Battery EOL usable capacity fraction & \(\fBatEOL\) & 0.7--0.9, set by cycles and temperature \\
Radiator emissivity & \(\epsrad\) & 0.8--0.95, surface and degradation dependent \\
Radiator view factor & \(\etaview\) & 0.7--0.9, set by pointing, blockage, and self-view \\
Radiator temperature & \(\Trad\) & 300--400 K, junction/transport/reliability limited \\
Effective radiator areal density & \(\sigRad\) & 2--10 kg/m\(^2\), including structure and puncture tolerance \cite{Juhasz1998Radiators,Gilmore2002Thermal} \\
Demonstrated optical downlink class & \(R_{\rm dl}\) & \(\sim100\)--200 Gb/s peak demonstrations \cite{Schieler2023TBIRD,Wang2025TBIRDOverview} \\
\bottomrule
\end{tabular}
\end{table*}

If the node must sustain $\Ptot$ continuously (including eclipse), and the sunlight duty fraction is $\etaSun$,
then the required PV electrical power during sunlit operation is obtained from an orbit-averaged energy balance:
\begin{equation}
P_{\mathrm{PV},\odot}\;\simeq\;\frac{\Ptot}{\etaPMAD}\,
\left[1+\frac{1-\etaSun}{\etaSun\,\eta_b}\right],
\label{eq:Ppv_sun}
\end{equation}
where $\eta_b$ is the storage round-trip efficiency. The bracketed term equals $1/\etaSun$ when $\eta_b\to 1$,
and exceeds $1/\etaSun$ when storage losses are included. This makes explicit that eclipse penalties appear twice: they increase both required stored energy (Sec.~\ref{sec:eclipse_storage}) and the PV charging burden.

Define the effective PV plane electrical power density (including cosine/packing/temperature margins)
as $q_{\rm PV}\equiv \etaPV\,\etaGeom\,\Ssol$. The required PV area is then
\begin{equation}
\Apv \;\simeq\; \frac{P_{\mathrm{PV},\odot}}{q_{\rm PV}}
\;=\;\frac{\Ptot}{\etaPMAD\,\etaPV\,\etaGeom\,\Ssol}\,
\left[1+\frac{1-\etaSun}{\etaSun\,\eta_b}\right].
\label{eq:Apv}
\end{equation}

For mass, we use the deployed PV \emph{system} specific power $\spPV$ (W/kg), which is the appropriate metric
because it includes blankets/cells, structural support, deployment, and harness. The PV mass is
\begin{equation}
M_{\mathrm{PV}} \;\simeq\; \frac{P_{\mathrm{PV},\odot}}{\spPV}
\;=\;\frac{\Ptot}{\etaPMAD\,\spPV}\,
\left[1+\frac{1-\etaSun}{\etaSun\,\eta_b}\right].
\label{eq:MPV}
\end{equation}

Normalizing by delivered IT power yields a refined PV contribution to mass-per-power:
\begin{equation}
\mkwPV\;\equiv\;\frac{M_{\mathrm{PV}}}{\Pit/\mathrm{kW}}
\;\simeq\;\frac{1000\,\alphaOH}{\etaPMAD\,\spPV}\,
\left[1+\frac{1-\etaSun}{\etaSun\,\eta_b}\right].
\label{eq:mkw_pv_refined}
\end{equation}
Eq.~(\ref{eq:mkw_pv_refined}) reduces to the simpler $\mkwPV\simeq 1000\alphaOH/(\etaSun\spPV)$ used in
first-pass sizing when $\etaPMAD\to 1$ and $\eta_b\to 1$; the refinement here makes the penalty terms explicit.
The first-order sensitivity remains \(\mkwPV\propto \spPV^{-1}\): doubling deployed PV specific power halves the PV contribution to mass-per-power, all else equal.

We define two compact penalty factors
\begin{align}
F_{\rm PV} &\equiv \frac{1}{\eta_{\rm PMAD}}
\left[1+\frac{1-f_\odot}{f_\odot \eta_b}\right]\frac{1}{f_{\rm EOL,PV}}, 
\label{eq:Fpv}\\
F_{\rm bat} &\equiv \frac{1}{\rho_{\rm DoD} f_{\rm EOL,bat}}
\label{eq:Fbat}.
\end{align}

\begin{figure*}[t]
\centering
\begin{minipage}{0.48\textwidth}
\centering
\begin{tikzpicture}
\begin{axis}[
width=0.95\linewidth,
height=0.58\linewidth,
xlabel={PV system specific power \(\spPV\) (W/kg)},
ylabel={\(\mkwPV\) (kg/kW)},
xmin=25,xmax=170,
ymin=0,ymax=85,
legend style={at={(0.65,0.96)},anchor=north,legend columns=1,draw=none,fill=none}
]
\addplot[domain=25:170,samples=250] {1000*1.25/(0.92*x)*(1+(1-0.95)/(0.95*0.90))};
\addlegendentry{high-sunlight, \(f_{\odot}=0.95\)}
\addplot[domain=25:170,samples=250,dashed] {1000*1.25/(0.92*x)*(1+(1-0.65)/(0.65*0.90))};
\addlegendentry{eclipsing, \(f_{\odot}=0.65\)}
\end{axis}
\end{tikzpicture}
\end{minipage}\hfill
\begin{minipage}{0.48\textwidth}
\centering
\begin{tikzpicture}
\begin{axis}[
width=0.95\linewidth,
height=0.58\linewidth,
xlabel={radiator temperature \(\Trad\) (K)},
ylabel={\(\Arad/\Ptot\) (m\(^2\)/MW)},
xmin=300,xmax=440,
ymin=500,ymax=5000,
ymode=log,
legend style={at={(0.70,0.96)},anchor=north,legend columns=1,draw=none,fill=none}
]
\addplot[domain=300:440,samples=240] {1e6/(0.90*0.85*5.670374419e-8*x^4)};
\addlegendentry{no parasitic load}
\addplot[domain=300:440,samples=240,dashed] {1e6/(0.90*0.85*5.670374419e-8*x^4 - 150)};
\addlegendentry{\(q_{\rm env}=150~\mathrm{W\,m^{-2}}\)}
\end{axis}
\end{tikzpicture}
\end{minipage}
\caption{First-order technology levers in the subsystem floor. The photovoltaic term decreases nearly as \(\spPV^{-1}\), while the radiator term is controlled by the nonlinear \(\Trad^{-4}\) dependence and the denominator penalty from absorbed environmental heat. Legends are placed outside the plotting region so that the curves remain unobscured.}
\label{fig:tech_levers}
\end{figure*}
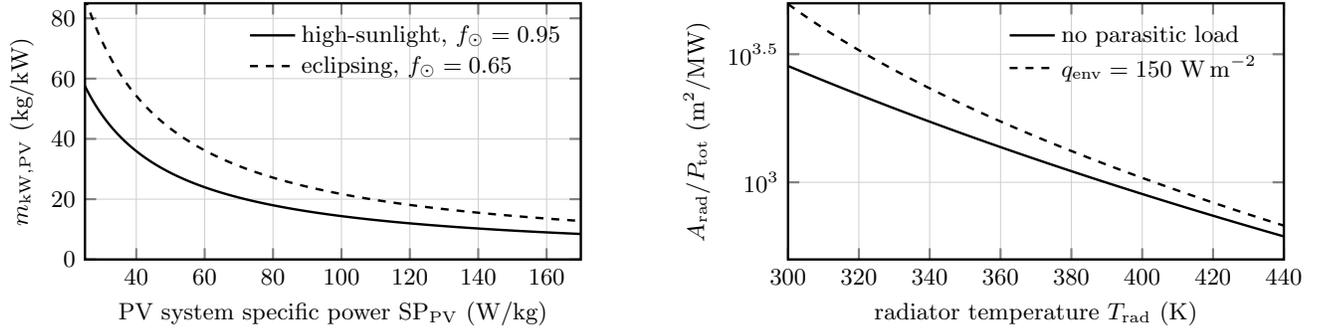

Figure~\ref{fig:tech_levers} highlights the two highest-leverage subsystem knobs in the present model. On the power side, deployed PV specific power drives \(\mkwPV\) almost exactly as \(\spPV^{-1}\). On the thermal side, allowable radiator temperature and absorbed environmental heat load control area through the denominator of Eq.~(\ref{eq:Arad_net}); this is why high-temperature-tolerant electronics and low-parasitic-flux geometries are not refinements but core technology levers.

\subsection{Eclipse storage: orbit geometry, usable capacity, and cycle-life penalties}
\label{sec:eclipse_storage}

Let $t_{\mathrm{ecl}}$ denote the eclipse duration per orbit and $\Torbit$ the orbital period. For a circular orbit of radius $r$ and solar beta angle $\betaang$ (angle between the Sun vector and the orbital plane), a standard umbra-geometry approximation gives the eclipse half-angle \cite{Vallado2013}
\begin{equation}
\phi_{\rm ecl} \;\simeq\;
\arccos\Big(\frac{\sqrt{1-(\Rae/r)^2}}{\cos\betaang}\Big),
\qquad
\betaang < \beta_{\rm crit}\equiv \arcsin(\Rae/r),
\label{eq:eclipse_halfangle}
\end{equation}
with $t_{\mathrm{ecl}} \simeq (\phi_{\rm ecl}/\pi)\Torbit$ and $t_{\mathrm{ecl}}=0$ when $\betaang\ge \beta_{\rm crit}$. This makes explicit that $\etaSun = 1-t_{\mathrm{ecl}}/\Torbit$ is an \emph{orbit-selection variable}, a system-level parameter. With orbital period $\Torbit $ given as usual 
\begin{equation}
\Torbit \;=\; 2\pi\sqrt{\frac{r^3}{\muE}},
\label{eq:Torb}
\end{equation}
where $\mu_\oplus \equiv G M_\oplus$ is Earth's standard gravitational parameter. Unless stated otherwise, we express $\Torbit$ and $t_{\rm ecl}$ in hours so that $\Ptot\,t_{\rm ecl}$ is in Wh when $\Ptot$ is in W, consistent with $e_b$ in Wh/kg. Representative orbit cases and their first-order implications are summarized later in Table~\ref{tab:orbit_rad_cases}.

The required delivered eclipse energy at the regulated bus is $E_{\rm ecl}=\Ptot\,t_{\mathrm{ecl}}$. Battery mass is set by \emph{usable} end-of-life capacity, not nameplate capacity. Let $e_b$ be cell-level specific energy (Wh/kg), $\rhoDoD$ the allowable depth-of-discharge fraction (typically chosen by cycle-life and reliability constraints), and $\fBatEOL$ the fraction of beginning-of-life capacity available at end-of-life under the expected cycle count and temperature. Then a capacity-driven sizing relation is
\begin{equation}
M_b \;\simeq\;
\frac{\Ptot\, t_{\mathrm{ecl}}}{(e_b\,\rhoDoD\,\fBatEOL)}.
\label{eq:Mbatt}
\end{equation}
Eq.~(\ref{eq:Mbatt}) isolates the \emph{mass} driver (usable capacity), while the \emph{energy} driver (how much PV charging is required) is carried separately by $\eta_b$ in Eq.~(\ref{eq:Ppv_sun}).
Finally, storage is also constrained by cycle throughput. If eclipse occurs each orbit, the number of charge/discharge
events over mission duration $T_{\rm miss}$ is $N_{\rm cyc}\sim T_{\rm miss}/\Torbit$. For \LEO\ this can reach
$\mathcal{O}(10^4)$--$\mathcal{O}(10^5)$ partial cycles over multi-year missions, which is why $\rhoDoD$ and $\fBatEOL$ cannot be treated as minor second-order details in economic sizing. 

\subsection{End-of-life sizing and the overhead factor as first-order design variables}
\label{sec:eol_overhead}

All power and mass relations should be interpreted as \emph{end-of-life} (EOL) constraints.
Rather than embedding lifetime penalties implicitly, it is cleaner to introduce explicit EOL factors.

\emph{PV end-of-life margin.} Let $\fPVEOL\in(0,1]$ denote the ratio of EOL PV output to beginning-of-life (BOL) PV output under the expected
radiation environment, contamination, thermal cycling, and pointing law.
Then the BOL PV sizing is obtained by multiplying Eqs.~(\ref{eq:Apv})--(\ref{eq:MPV}) by $1/\fPVEOL$:
\begin{equation}
\Apv^{\rm BOL}\simeq \frac{\Apv^{\rm EOL}}{\fPVEOL},
\qquad
M_{\rm PV}^{\rm BOL}\simeq \frac{M_{\rm PV}^{\rm EOL}}{\fPVEOL}.
\label{eq:PV_EOL_factor}
\end{equation}
This isolates the design question: what $\fPVEOL$ is credible for the chosen orbit and lifetime?

\emph{Battery end-of-life margin.} Battery fade enters through $\fBatEOL$ in Eq.~(\ref{eq:Mbatt}) and is primarily a function of cycle count,
temperature, and allowable $\rhoDoD$. Because $\fBatEOL$ is mission- and orbit-dependent, it should be treated
explicitly (rather than buried inside $e_b$).

\emph{Overhead factor and electrical efficiency accounting.} To separate conversion losses from housekeeping loads, write
\begin{equation}
\Ptot \;=\; \frac{\Pit}{\eta_{\rm dc}} \;+\; P_{\rm hk},
\qquad
\alphaOH \;=\; \frac{\Ptot}{\Pit} \;=\; \frac{1}{\eta_{\rm dc}}+\frac{P_{\rm hk}}{\Pit},
\label{eq:alpha_split}
\end{equation}
where $\eta_{\rm dc}$ is the effective electrical efficiency from regulated bus to IT loads (VRMs, distribution),
and $P_{\rm hk}$ is aggregate housekeeping power (thermal transport, ADCS/pointing, comm terminals, propulsion averaged over time, avionics). A practical further decomposition is
\begin{equation}
\frac{P_{\rm hk}}{\Pit}=\alpha_{\rm th}+\alpha_{\rm com}+\alpha_{\rm adcs}+\alpha_{\rm prop}+\cdots,
\label{eq:alpha_budget_refined}
\end{equation}
which makes $\alphaOH$ experimentally and operationally verifiable. Once a platform architecture is specified, $\alphaOH$ enters the first-order PV, storage, and radiator sizing relations through $P_{\rm tot}=\alpha_{\rm OH}P_{\rm IT}$; however, the subsystems that determine $\alphaOH$ need not scale linearly with node power. Communications terminal power, thermal-transport power, pointing control, and collision-avoidance activity can all change nonlinearly with throughput, aperture, radiator layout, and orbit. The key architectural point is that $\alphaOH$ is a primary sizing variable: underestimating it directly underestimates PV mass, radiator area, storage recharge burden, and the mass-per-power threshold in the break-even equations.

Although the payload is abstracted behind $\Pit$, the abstraction is not architecture-neutral. Memory capacity, memory bandwidth, interconnect topology, and checkpointing doctrine feed directly into both $\alphaOH$ and $\Gamma$. High-radix onboard fabrics, longer intra-cluster data paths, and stronger fault-tolerance increase housekeeping power through switching, transceiver, thermal-transport, and synchronization overhead; disaggregated memory, parameter synchronization, and checkpoint-heavy workloads can increase the space-to-ground fraction $f_{\rm ext}$ and thereby tighten the condition $\Gamma_{\rm ext}\le \Gamma_{\max}$. In the present model these effects enter through $\alphaOH$, $U_{\rm eff}$, and $\Gamma_{\rm ext}$ rather than through a bottom-up compute-architecture model.

\section{Thermal rejection: radiative area, temperature margin, and puncture tolerance}
\label{sec:thermal}

In vacuum there is no convective heat sink. Steady-state heat rejection is therefore dominated by radiation from deployable surfaces. For compute payloads, nearly all electrical power drawn by the platform is ultimately dissipated as heat that must be transported from semiconductor junctions to cold plates, through a working fluid or conductive path, and finally to deployed radiators. Thermal closure is therefore a geometry, mass, reliability, and operations constraint, not a secondary packaging detail.

\subsection{Radiator area scales as \(T^{-4}\): the idealized bound}

Let $\dot Q_{\rm rej}$ be the net heat that must be rejected by the thermal system. For a compute-dominated node,
optical power exported by laser communications and other non-thermal outputs is typically negligible relative to $\Ptot$ at the system level, so to first order
\begin{equation}
\dot Q_{\rm rej} \simeq \Ptot = \alphaOH \Pit.
\end{equation}
In the idealized case where the radiator sees deep space with an effective view factor $\etaview$ and has effective
emissivity $\epsrad$, the Stefan--Boltzmann relation yields
\begin{equation}
\Ptot \simeq \epsrad\,\etaview\,\sigSB\,\Arad\,\Trad^4,
\label{eq:rad}
\end{equation}
and therefore
\begin{equation}
\Arad \simeq \frac{\Ptot}{\epsrad\,\etaview\,\sigSB\,\Trad^4}.
\label{eq:Arad}
\end{equation}
The scaling $\Arad \propto \Trad^{-4}$ is the key result for the idealized sink. A temperature increase from 350 to 385 K, for example, reduces the ideal radiator area by a factor $1.1^4\simeq1.46$, provided the semiconductor junction-temperature and transport-margin constraints in Sec.~\ref{sec:thermal_transport} remain satisfied. The local logarithmic sensitivity is
\begin{equation}
\frac{d\ln \Arad}{d\ln \Trad} = -4,
\end{equation}
so a 10\% increase in $\Trad$ reduces $\Arad$ by $\approx 1.1^4\simeq 1.46\times$. 

For a concrete anchor, the emitted heat flux (ignoring environment) is
\begin{equation}
q_{\rm emit} \equiv \epsrad\,\etaview\,\sigSB\,\Trad^4
\approx \SI{651}{W\,m^{-2}}
\left(\frac{\epsrad}{0.90}\right)\left(\frac{\etaview}{0.85}\right)
\left(\frac{\Trad}{350~\mathrm{K}}\right)^4,
\end{equation}
so a \SI{1}{MW} rejected-heat requirement implies $\Arad\sim \SI{1.5e3}{m^2}$ at \SI{350}{K} even before
parasitic environmental heating is included.

\subsection{Net radiator performance in Earth orbit: environmental heat loads}

Eq.~(\ref{eq:rad}) is an optimistic upper bound because Earth-orbit radiators absorb time-varying direct solar, albedo, infrared, and spacecraft-self-view heat fluxes. A compact net form is
\begin{equation}
\Ptot \simeq \Arad\left(\epsrad\,\etaview\,\sigSB\,\Trad^4 - q_{\mathrm{env}}\right),
\label{eq:rad_net}
\end{equation}
where $q_{\mathrm{env}}$ is the effective absorbed environmental heat flux averaged over the radiator pointing law.
Solving,
\begin{equation}
\Arad \simeq
\frac{\Ptot}{\epsrad\,\etaview\,\sigSB\,\Trad^4 - q_{\mathrm{env}}}.
\label{eq:Arad_net}
\end{equation}
This form is the architecture-level thermal closure condition: absorbed environmental heat reduces the net radiative heat flux available for payload waste heat. If $q_{\rm env}$ is a
fraction $\xi$ of the ideal emitted flux ($q_{\rm env}=\xi\,q_{\rm emit}$), then $\Arad$ is inflated by
\begin{equation}
\Arad = \frac{\Arad|_{q_{\rm env}=0}}{1-\xi}.
\end{equation}
Thus even $\xi=0.2$ increases area by 25\%, and $\xi=0.4$ increases area by 67\%.

A geometry-resolved decomposition for $q_{\rm env}$ is
\begin{equation}
q_{\mathrm{env}} \;\approx\;
\alpha_{\rm sol}\,S_0\,f_{\odot}^{(\rm rad)}
+\alpha_{\rm sol}\,A_{\rm B}\,S_0\,f_{\rm alb}
+\epsrad\,\sigma_{\rm SB}\,T_{\oplus}^{4}\,f_{\rm IR}
+q_{\rm self},
\label{eq:qenv_decomp}
\end{equation}
where $\alpha_{\rm sol}$ is solar absorptivity of the radiator surface, $A_{\rm B}$ is Earth Bond albedo,
$f_{\odot}^{(\rm rad)}$ is the radiator's time-averaged effective exposure to direct sun, $f_{\rm alb}$ and $f_{\rm IR}$
encode the Earth albedo and Earth IR geometric couplings (view factors and pointing), and $q_{\rm self}$ captures
spacecraft self-view and scattered light. A mission design may equivalently express the net radiative term as $\epsilon_{\rm rad}\eta_{\rm view}\sigma_{\rm SB}\big(T_{\rm rad}^{4}-T_{\rm sink,eff}^{4}\big)$, where $T_{\rm sink,eff}$ is an effective environment temperature inferred from $q_{\rm env}$. For MW-class nodes, thermal closure requires a radiator pointing law and sunshielding approach that drives $f_{\odot}^{(\rm rad)}\!\rightarrow\!0$ and keeps $f_{\rm IR},f_{\rm alb}$ small; otherwise the denominator in Eq.~(\ref{eq:Arad_net}) decreases and required area increases rapidly \cite{Gilmore2002Thermal,Juhasz1998Radiators}.

The operative radiator sizing used throughout is Eq.~(\ref{eq:Arad_net}), not a linearization. The only local linearized statement in this section is the logarithmic sensitivity \(d\ln \Arad/d\ln \Trad=-4\), which is used solely to expose leverage around a chosen operating point. The lumped form in Eq.~(\ref{eq:Arad_net}) is intended for architecture-level sizing when (i) a radiator can be represented by a single effective temperature \(\Trad\), (ii) the environment can be absorbed into a time-averaged \(q_{\rm env}\), and (iii) segmentation, headers, manifolds, and puncture tolerance are represented through an effective areal density \(\sigma_{\rm rad}\). It is not a substitute for panel-resolved thermal analysis, transient shadowing analysis, or a detailed fluid-network model.
\subsection{Radiator mass and puncture tolerance: effective areal density}

With radiator areal density $\sigRad$ (kg/m$^2$), radiator mass scales linearly with area,
\begin{equation}
M_{\rm rad} \simeq \sigRad\,\Arad.
\end{equation}
For MW-class nodes, Eq.~(\ref{eq:Arad_net}) already implies multi-ton radiators even for modest $\sigRad$. Importantly, $\sigRad$ should be interpreted as an \emph{effective} areal density that already includes manifold/headers, deployment structure, micrometeoroid/debris puncture tolerance, and (often) redundancy required to avoid single-point loss of coolant loops \cite{Juhasz1998Radiators,Gilmore2002Thermal}.
Radiator areal density is therefore a system-level quantity, not only a panel-material property.

A compact reliability scaling that motivates segmentation is the Poisson penetration model. Let $\phi_p$ be the
penetration rate (m$^{-2}$yr$^{-1}$) for the relevant puncture-causing particle population under the assumed orbit, shielding, and radiator technology \cite{NASAORDEM,NASASTD8719_14B}. For an exposed area $\Arad$ and mission duration $T$, the expected penetrations are
$\mu=\phi_p\,\Arad\,T$ and the probability of \emph{zero} penetrations is
\begin{equation}
P_0=\exp(-\phi_p\,\Arad\,T).
\label{eq:puncture_poisson}
\end{equation}
Because $\mu \propto \Arad$, MW-class radiators require segmentation, isolation, and graceful degradation rather than reliance on a single unsegmented heat-rejection loop.
Those mitigations increase $\sigRad$ and increase $\mkw[\mathrm{fixed}]$ through valves, control, sensing, and structural duplication.

\subsection{Transport from chips to radiators: junction-temperature and overhead-power couplings}
\label{sec:thermal_transport}

Radiator temperature $\Trad$ is not freely selectable: it is upper-bounded by allowable junction temperature $T_j$
and by the total thermal resistance from junction to radiator. A thermal-resistance budget is
\begin{equation}
T_j \;=\; \Trad \;+\;\Delta T_{\rm pkg} \;+\;\Delta T_{\rm coldplate}\;+\;\Delta T_{\rm loop}\;+\;\Delta T_{\rm rad},
\label{eq:Tj_budget}
\end{equation}
where the $\Delta T$ terms include packaging/heat-spreader drops, cold-plate/interface drops, loop transport drop,
and radiator internal conduction drop. A condensed resistance representation is
\begin{equation}
T_j \simeq \Trad + \Ptot\,R_{\theta},
\label{eq:Tj}
\end{equation}
with $R_{\theta}$ the effective junction-to-radiator thermal resistance. The $T^{-4}$ benefit of raising $\Trad$ therefore competes directly with semiconductor junction-temperature limits, leakage power, timing margin, and thermal-throttling thresholds. A radiator operating point near 350 K is already close to 77\,$^\circ$C before package, cold-plate, loop, and panel conduction drops are added; high-performance processors generally require substantially lower junction temperature than the maximum destructive limit to maintain efficiency and lifetime.

Moreover, achieving low $R_{\theta}$ and controlling $\Delta T_{\rm loop}$ can impose non-negligible pumping power.
For a single-phase loop with coolant specific heat $c_p$, the required mass flow to transport heat is
$\dot m \simeq \Ptot/(c_p\,\Delta T_{\rm loop})$, and the pump power is approximately
\begin{equation}
P_{\rm pump} \simeq \frac{\Delta p}{\eta_p}\frac{\dot m}{\rho},
\label{eq:pump_power}
\end{equation}
where $\Delta p$ is pressure drop, $\eta_p$ pump efficiency, and $\rho$ fluid density. Since $\Delta p$ typically increases with flow, reducing $\Delta T_{\rm loop}$ to preserve junction margin can increase $P_{\rm pump}$ superlinearly, feeding the overhead factor $\alphaOH$ introduced in Eq.~(\ref{eq:Ptot_def}). Two-phase transport can reduce mass flow via latent heat, but generally increases system complexity and fault-management requirements; these penalties must be carried in $\alphaOH$ and in $\mkw[\mathrm{fixed}]$.

In summary, the thermal subsystem is a first-order architecture driver. It sets deployed span and inertia, exposed area, a major fraction of mass-per-power through \(\sigRad\Arad\), and a nontrivial component of overhead power through pumping and control. This is why MW-class orbital compute tends to reside in the tens-of-kg/kW regime even under optimistic assumptions and why high-temperature-tolerant electronics, low-parasitic-flux geometries, and puncture-tolerant radiator architectures are leverage points rather than refinements.

\section{A compact mass-per-power decomposition}
\label{sec:mkw}

For break-even economics we expose \mkw\ as the sum of identifiable, physically interpretable
subsystem contributions. Throughout, \Pit\ and \Ptot\ are treated as \emph{watts}; \mkw\ is reported in
\si{kg/kW}, hence the explicit factor \(1000\) that converts \si{kg/W} to \si{kg/kW}.

Sections~\ref{sec:power} and \ref{sec:thermal} establish the architecture-invariant subsystem floor. The purpose of the present section is to separate that floor from the architecture-sensitive choices—especially node granularity and gateway specialization—that determine how it maps into constellation-level competitiveness.

Starting from the PV mass relation Eq.~(\ref{eq:MPV}), the eclipse-storage mass Eq.~(\ref{eq:Mbatt}),
and the radiator requirement [Eq.~(\ref{eq:Arad}) together with \(M_{\rm rad}\simeq \sigRad \Arad\)],
and substituting the overhead accounting \(\Ptot=\alphaOH\Pit\) [Eq.~(\ref{eq:Ptot_def})], we obtain

\begin{equation}
\mkw \;\approx\;
\underbrace{
\frac{1000\,\alphaOH}{\etaPMAD\,\spPV}
\left[1+\frac{1-\etaSun}{\etaSun\,\eta_b}\right]\frac{1}{\fPVEOL}
}_{\mkwPV}
+
\underbrace{
\frac{1000\,\alphaOH\,t_{\rm ecl}}{e_b\,\rhoDoD\,\fBatEOL}
}_{\mkwbat}
+
\underbrace{
\frac{1000\,\alphaOH\,\sigRad}{\epsrad\,\etaview\,\sigSB\,\Trad^4-q_{\rm env}}
}_{\mkwrad}
+\mkwfixed.
\label{eq:mkw_decomp}
\end{equation}
where \(\mkwfixed\) captures structure, deployment hardware, propulsion/propellant, avionics/ADCS,
comms terminals, shielding margin, harness, and compute packaging. Eq.~(\ref{eq:mkw_decomp})
is the central constraint equation: it states explicitly which levers reduce orbital mass-per-power
(\(\spPV,\Trad,\etaSun\)) and which penalties (eclipse \(t_{\rm ecl}\), view factor \(\etaview\), puncture tolerance via \(\sigRad\))
push it upward.

The corresponding \emph{physics floor}
\begin{equation}
\mkw \;\ge\; \mkwPV+\mkwbat+\mkwrad,
\label{eq:mkw_floor}
\end{equation}
since \(\mkwfixed\ge 0\). This makes clear that cost reductions in launch alone cannot compensate for unfavorable
\(\spPV\), \(\Trad\), or \(t_{\rm ecl}\): the platform must move \emph{both} \((\Lkg+\Bkg)\) and \mkw\ favorably.

Finally, Eq.~(\ref{eq:mkw_decomp}) provides immediate parametric sensitivities that identify high-leverage technology knobs:
\begin{equation}
\frac{\partial \ln \mkwPV}{\partial \ln \spPV}=-1,\qquad
\frac{\partial \ln \mkwrad}{\partial \ln \Trad}=-4,\qquad
\frac{\partial \ln \mkwdot}{\partial \ln \alphaOH}=+1,
\label{eq:mkw_sens}
\end{equation}
i.e., doubling \(\spPV\) halves \(\mkwPV\), while increasing \(\Trad\) from 350~K to 400~K reduces \(\mkwrad\) by a factor
\((350/400)^4\simeq 0.59\) (provided the junction-to-radiator thermal lift and reliability constraints can tolerate it).

\subsection{Bounding $\mkwfixed$ via two reference architectures}
\label{sec:mkw_fixed_arch}

The term $\mkwfixed$ in Eq.~(\ref{eq:mkw_decomp}) aggregates all spacecraft and payload mass not already
captured by the PV, battery, and radiator physics terms. For sizing and economic sensitivity we
to decompose it into budget lines that map to standard spacecraft subsystems:
\begin{equation}
M_{\rm fixed} \equiv
M_{\rm str} + M_{\rm adcs} + M_{\rm av} + M_{\rm prop} + M_{\rm com} + M_{\rm pkg} + M_{\rm sh} + M_{\rm marg},
\label{eq:Mfixed_decomp}
\end{equation}
where $M_{\rm str}$ is structure/deployment, $M_{\rm adcs}$ is attitude determination and control,
$M_{\rm av}$ is avionics and harness, $M_{\rm prop}$ is propulsion plus propellant (including end-of-life disposal reserve),
$M_{\rm com}$ is the communications payload (terminals, gimbals, modems),
$M_{\rm pkg}$ is compute packaging/coldplates/power distribution local to IT loads,
$M_{\rm sh}$ is shielding margin, and $M_{\rm marg}$ is growth and qualification margin.

Consistent with the convention used in Eq.~(\ref{eq:mkw_decomp}) (with $\Pit$ in W), we define
\begin{equation}
\mkwfixed \equiv \frac{1000\,M_{\rm fixed}}{\Pit}\qquad [\mathrm{kg/kW}].
\label{eq:mkw_fixed_def}
\end{equation}

\begin{table*}[t]
\caption{Fixed-mass budget ranges for two reference architectures. The ranges bound \(\mkwfixed\) and expose the main non-PV, non-storage, and non-radiator contributors. Optical-terminal mass can be in the few-kilogram class for \(\mathcal{O}(10^2)\) Gb/s demonstrations \cite{Schieler2023TBIRD}, but operational systems add redundancy, gimbals, and margin.}
\label{tab:mkw_fixed_budgets}
\centering
\small
\setlength{\tabcolsep}{5pt}
\renewcommand{\arraystretch}{1.12}
\begin{tabular}{@{}l r r@{}}
\toprule
Subsystem line item & \parbox[t]{0.22\textwidth}{\centering \(\Pit=\SI{1}{MW}\) monolithic node [kg]} & \parbox[t]{0.24\textwidth}{\centering \(\Pitn=\SI{100}{kW}\) modular node [kg/node]} \\
\midrule
Structure and deployment \(M_{\rm str}\) & 2,000--6,000 & 200--700 \\
ADCS/GNC \(M_{\rm adcs}\) & 200--900 & 30--120 \\
Avionics and harness \(M_{\rm av}\) & 150--600 & 20--80 \\
Propulsion and propellant \(M_{\rm prop}\) & 800--4,000 & 80--500 \\
Communications payload \(M_{\rm com}\) & 20--300 & 5--80 \\
Compute packaging/thermal transport \(M_{\rm pkg}\) & 3,000--12,000 & 300--2,000 \\
Shielding margin \(M_{\rm sh}\) & 200--3,000 & 20--400 \\
Growth and qualification margin \(M_{\rm marg}\) & 10--25\% of subtotal & 10--25\% of subtotal \\
\midrule
Implied \(M_{\rm fixed}\) & \(\sim7{,}000\)--\(30{,}000\) & \(\sim700\)--\(4{,}000\) \\
Implied \(\mkwfixed\) [Eq.~(\ref{eq:mkw_fixed_def})] & \(\sim7\)--30 kg/kW & \(\sim7\)--40 kg/kW \\
\bottomrule
\end{tabular}
\end{table*}

Table~\ref{tab:mkw_fixed_budgets} provides two reference architectures---a monolithic MW-class node and a modular distributed architecture---to bound representative $\mkwfixed$ and to make the scaling assumptions explicit in a spacecraft-systems sense (structures, GNC, propulsion, comm, packaging, and margins).

In the remainder of the paper, the distributed constellation is the baseline architectural interpretation. The monolithic \(\Pit=\SI{1}{MW}\) case is retained because it cleanly exposes the subsystem physics floor, but it is a subsystem-physics anchor rather than the baseline deployment concept.
\subsection{Constellation-first architecture: node granularity, gatewaying, and shared mass}
\label{sec:constellation}

The operationally relevant orbital-compute deployment is not a single MW-class spacecraft but a distributed constellation of smaller compute nodes linked by optical inter-satellite links (ISLs) and serviced by a smaller number of gateway nodes. This architecture is motivated both by resilience and by current public system concepts, which emphasize modularity, ISLs, high-sunlight LEO operation, and increasingly explicit gateway-layer assumptions \cite{GoogleSuncatcherPaper2025,ESPISBDC2025,FCCSpaceXODC2026,NVIDIASpaceComputing2026}.

\begin{table*}[t]
\caption{Anchored distributed reference cluster used to give \(\mkwfixed\) a concrete scale. The values are not optimized; they are a mid-band realization consistent with the modular range discussed in the text. With the high-sunlight base-case subsystem assumptions used in the worked anchor, the fixed-mass subtotal implies \(\mkw^{(\Sigma)}\approx49~\mathrm{kg/kW}\) before reserve margin.}
\label{tab:ref_cluster_nominal}
\centering
\small
\setlength{\tabcolsep}{4pt}
\renewcommand{\arraystretch}{1.14}
\begin{tabular}{@{}l c l@{}}
\toprule
Parameter & Nominal value & Interpretation \\
\midrule
Orbit shell & dawn-dusk \LEO & \parbox[t]{0.43\textwidth}{\raggedright High-sunlight operating point that suppresses the battery term in first-pass economics.} \\
Compute-node power \(\Pitn\) & \(\SI{100}{kW}\) & \parbox[t]{0.43\textwidth}{\raggedright Mid-band value for the distributed baseline.} \\
Active compute nodes \(N_c\) & 20 & \parbox[t]{0.43\textwidth}{\raggedright Aggregate cluster IT power \(P_{\mathrm{IT},\Sigma}=\SI{2}{MW}\).} \\
Gateway nodes \(N_{\rm gw}\) & 4 & \parbox[t]{0.43\textwidth}{\raggedright One gateway for every five compute nodes; concentrates space-to-ground traffic at cluster level.} \\
Local fixed mass per compute node \(M_{\rm loc}\) & \(\SI{1.5}{t}\) & \parbox[t]{0.43\textwidth}{\raggedright Local fixed-mass penalty \(1000M_{\rm loc}/\Pitn\approx\SI{15}{kg/kW}\).} \\
Cluster-shared gateway/network mass \(M_{\rm sh}\) & \(\SI{10}{t}\) total & \parbox[t]{0.43\textwidth}{\raggedright Shared penalty \(1000M_{\rm sh}/P_{\mathrm{IT},\Sigma}\approx\SI{5}{kg/kW}\).} \\
Illustrative fixed subtotal & \(\SI{20}{kg/kW}\) & \parbox[t]{0.43\textwidth}{\raggedright Explicit anchor inside the broader modular range; used for the break-even discussion.} \\
\noalign{\vskip 2pt}
\bottomrule
\end{tabular}
\end{table*}

Let \(N_c\) denote the number of compute nodes, each delivering \(\Pitn\), so that the total cluster IT power is
\begin{equation}
P_{\mathrm{IT},\Sigma}=N_c\,\Pitn.
\label{eq:Pcluster}
\end{equation}
Decompose non-PV, non-battery, non-radiator mass into a local per-node fixed term \(M_{\rm loc}\) and a cluster-shared term \(M_{\rm sh}\) (e.g., gateway payloads, shared reserves, formation-metrology overhead, and network-control hardware). Then the cluster-average mass-per-power is
\begin{equation}
\mkw^{(\Sigma)}
\;\approx\;
\mkwPV
+\mkwbat
+\mkwrad
+\frac{1000\,M_{\rm loc}}{\Pitn}
+\frac{1000\,M_{\rm sh}}{P_{\mathrm{IT},\Sigma}}.
\label{eq:mkw_cluster}
\end{equation}
Eq.~(\ref{eq:mkw_cluster}) makes the node-granularity trade explicit: smaller nodes increase local fixed-mass duplication through \(M_{\rm loc}/\Pitn\), while larger nodes reduce duplication but increase the capacity fraction lost per node failure,
\begin{equation}
\phi_1\equiv \frac{\Pitn}{P_{\mathrm{IT},\Sigma}}=\frac{1}{N_c}.
\label{eq:phi1}
\end{equation}
The relevant architectural question is where node power should lie between duplication-dominated and resilience-dominated regimes.

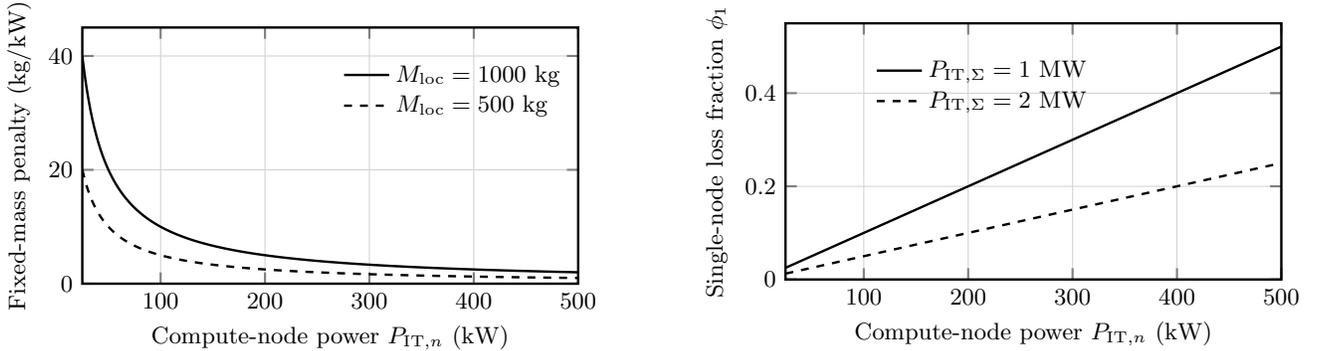
\begin{figure*}[t]
\centering
\begin{minipage}{0.48\textwidth}
\centering
\begin{tikzpicture}
\begin{axis}[
width=0.95\linewidth,
height=0.58\linewidth,
xlabel={Compute-node power \(\Pitn\) (kW)},
ylabel={Fixed-mass penalty (kg/kW)},
xmin=25,xmax=500,
ymin=0,ymax=45,
legend style={at={(0.75,0.90)},anchor=north,legend columns=1,draw=none,fill=none}
]
\addplot[domain=25:500,samples=250] {1000/x};
\addlegendentry{\(M_{\rm loc}=1000~\mathrm{kg}\)}
\addplot[domain=25:500,samples=250,dashed] {500/x};
\addlegendentry{\(M_{\rm loc}=500~\mathrm{kg}\)}
\end{axis}
\end{tikzpicture}
\end{minipage}\hfill
\begin{minipage}{0.48\textwidth}
\centering
\begin{tikzpicture}
\begin{axis}[
width=0.95\linewidth,
height=0.58\linewidth,
xlabel={Compute-node power \(\Pitn\) (kW)},
ylabel={Single-node loss fraction \(\phi_1\)},
xmin=25,xmax=500,
ymin=0,ymax=0.55,
legend style={at={(0.4,0.90)},anchor=north,legend columns=1,draw=none,fill=none}
]
\addplot[domain=25:500,samples=2] {x/1000};
\addlegendentry{\(P_{\mathrm{IT},\Sigma}=1~\mathrm{MW}\)}
\addplot[domain=25:500,samples=2,dashed] {x/2000};
\addlegendentry{\(P_{\mathrm{IT},\Sigma}=2~\mathrm{MW}\)}
\end{axis}
\end{tikzpicture}
\end{minipage}
\caption{Constellation granularity trade for fixed total cluster power. Reducing node power lowers the fraction of capacity lost in one node failure but increases repeated local fixed mass; increasing node power reduces duplication but increases the service impact of a node loss. The selected node power is therefore set by the joint trade among fixed-mass duplication, reserve fraction, replacement cadence, gateway scaling, and service continuity.}
\label{fig:constellation_trade}
\end{figure*}

Gateway specialization changes the external communications constraint. If \(N_{\rm gw}\) gateway nodes each support \(N_{\rm term}\) space-to-ground terminals, then the aggregate cluster throughput is bounded by
\begin{equation}
\overline{R}_{\mathrm{sg},\Sigma}
\;\le\;
N_{\rm gw}\,N_{\rm term}\,f_{\rm link}\,R_{\rm pk},
\label{eq:Rcluster}
\end{equation}
and the corresponding space-to-ground link communication-intensity ceiling is
\begin{equation}
\Gamma_{\max}^{(\Sigma)}
\equiv
\frac{\overline{R}_{\mathrm{sg},\Sigma}}{\Util\,P_{\mathrm{IT},\Sigma}}.
\label{eq:Gamma_cluster}
\end{equation}
Eq.~(\ref{eq:Gamma_cluster}) shows why constellations help but do not eliminate the C3 bottleneck: internal ISLs and gateway nodes can relax per-node external I/O, but terrestrial-user workloads remain limited by aggregate space-to-ground throughput.

Service continuity also improves with distribution. If service requires at least \(k\) of \(N_c\) compute nodes and node availability is \(a\), then under an independence approximation
\begin{equation}
A_{\rm svc}=\sum_{i=k}^{N_c}\binom{N_c}{i}a^i(1-a)^{N_c-i}.
\label{eq:Asvc_cluster}
\end{equation}
This benefit must be balanced against correlated fault domains, which are treated later through \(\lambda_{\rm eff}\) in Sec.~\ref{sec:security}. For constellation architectures, compute nodes and gateway nodes should be distinguished explicitly. Internal ISLs can support cluster-local task distribution and state movement, while a smaller number of gateway nodes concentrate the space-to-ground link. This does not remove the external C3 bottleneck, but it changes its scaling from per-node to per-cluster, as in Eqs.~(\ref{eq:Rcluster})--(\ref{eq:Gamma_cluster}).

\subsection{Worked example: a \texorpdfstring{\(\Pit=\SI{1}{MW}\)}{1 MW} node (physics-floor areas and masses)}
\label{sec:worked_example}

The \(\Pit=\SI{1}{MW}\) node is retained only as a transparent subsystem-physics anchor. Its role is to show that the unavoidable PV+radiator subtotal already resides in the \(\mathcal{O}(20)\,\mathrm{kg/kW}\) regime before fixed spacecraft mass, gateway specialization, reserve policy, or space-to-ground link allocation are added. The full numerical treatment is therefore deferred to Sec.~\ref{sec:worked_structure}, where the same anchor is evaluated consistently as a physics-floor/base-case pair in Table~\ref{tab:worked_1mw_anchor}.

At MW scale, the remaining C3 question is whether fixed spacecraft mass, space-to-ground link scaling, and delivered-compute penalties can remain below the margin opened by that subsystem floor. Auxiliary area- and mass-scaling relations for this anchor are retained in the optional supplemental material. Keeping them out of the main text avoids a duplicate worked example without changing the paper's quantitative anchors.


\section{Communications: optical networking and ground availability}
\label{sec:comms}

For terrestrial-user orbital compute (C3), communications is not an implementation detail: it is the mechanism by which
orbit pays an unavoidable \emph{I/O and availability constraint} relative to terrestrial data centers. Communications enters
the feasibility problem in three coupled ways. First, it imposes a hard ceiling on sustained space-to-ground throughput
through geometry and weather duty factor (Sec.~\ref{sec:comms_sanity}). Second, achieving high availability pushes
toward a geographically diverse optical ground segment whose capital and operating cost scales with required aggregate throughput.
Third, the spacecraft communications payload (terminals, gimbals, fine pointing, coding/processing) consumes nontrivial mass and
electrical power, feeding directly into $\alphaOH$ [Eq.~(\ref{eq:Ptot_def})] and $\mkw[\mathrm{fixed}]$ [Eq.~(\ref{eq:mkw_decomp})].
We quantify this coupling using the communication-intensity metric $\Gamma$ defined in Sec.~\ref{sec:system}.

\subsection{Average throughput is duty-cycle limited}
\label{sec:comms_duty}

Optical downlink demonstrations in \LEO\ have achieved peak physical-layer rates in the 100--200~Gb/s class
under favorable conditions \cite{Hemmati2006,Schieler2023TBIRD,Wang2025TBIRDOverview,LCRD2023FactSheet}. The binding question for
C3 is not peak rate; it is \emph{sustained time-average} rate after accounting for (i) geometric access, (ii) weather,
(iii) protocol/pointing overhead, and (iv) ground-station scheduling when many nodes share finite ground resources.

Let $R_{\mathrm{pk}}$ be the peak physical-layer rate of a single optical terminal during usable contact.
Let $N_{\rm term}$ be the number of \emph{simultaneously usable} space-to-ground terminals per node (often 1 in early designs,
but a scalable architecture may require $N_{\rm term}>1$). Define an effective link duty factor $f_{\mathrm{link}}$ capturing
all contact and availability constraints. Then the sustained throughput is bounded by
\begin{equation}
\overline{R}_{\mathrm{sg}} \;\le\; N_{\rm term}\, f_{\mathrm{link}}\,R_{\mathrm{pk}}.
\label{eq:Ravg}
\end{equation}
For optical downlink, decompose
\begin{equation}
f_{\mathrm{link}} \;=\; \eta_{\rm link}\, f_{\rm vis}\, p_{\rm avail}\, f_{\rm sched},
\label{eq:flink_factored}
\end{equation}
where: (i) $f_{\rm vis}$ is the geometric visibility fraction (line-of-sight above an elevation mask to at least one site),
(ii) $p_{\rm avail}$ is the probability that at least one candidate optical site is cloud-free (or an RF fallback is available), (iii) $\eta_{\rm link}$ captures coding, reacquisition, fades, and protocol overhead, and (iv) $f_{\rm sched}$ accounts for ground network contention (time-division scheduling) when many satellites share a finite number of ground terminals.

At constellation scale, scheduling enters the duty factor explicitly. Let \(N_{\rm beam}\) denote the number of simultaneous optical links that each ground site can support, and let \(N_{\rm sat,act}\) denote the number of satellites requiring service over the same interval. The scheduling factor \(f_{\rm sched}\in(0,1]\) is then bounded by
\begin{equation}
f_{\rm sched}\;\lesssim\;\min\!\left(1,\ \frac{N_{\rm gs}\,N_{\rm beam}}{N_{\rm sat,act}}\right),
\label{eq:fsched}
\end{equation}
which shows that per-node sustained throughput degrades unless the ground segment scales approximately with offered load. Setting \(f_{\rm sched}=1\) is therefore a substantive assumption, not a neutral default.

Weather availability is similarly not a fixed constant. If each site has clear-sky probability \(p_{\rm cs}\) and the sites are treated as statistically independent, the optimistic availability is
\begin{equation}
p_{\rm avail} \;=\; 1-(1-p_{\rm cs})^{N_{\rm gs}}.
\label{eq:pavail2}
\end{equation}
Cloud fields are often correlated over regional scales, so it is conservative to introduce an effective independent-site count
\begin{equation}
N_{\rm eff}\;\equiv\;\frac{N_{\rm gs}}{1+\rho_w(N_{\rm gs}-1)},\qquad
p_{\rm avail}\;\approx\;1-(1-p_{\rm cs})^{N_{\rm eff}},
\label{eq:pavail_corr}
\end{equation}
where \(\rho_w\in[0,1)\) is a weather-correlation parameter. Eq.~(\ref{eq:pavail2}) should therefore be read as an upper bound, with Eq.~(\ref{eq:pavail_corr}) providing a more conservative planning approximation.

These relations also close the feedback between communications and spacecraft mass. Increasing peak rate \(R_{\rm pk}\), terminal count \(N_{\rm term}\), or pointing-stability requirement raises terminal mass and electrical overhead, thereby increasing \(\alphaOH\) and \mkwfixed. Communications relief is therefore never free in the mass or cost model.

\subsection{Representative throughput and communication ceiling}
\label{sec:comms_sanity}

A minimal throughput model follows directly from Eqs.~(\ref{eq:Ravg})--(\ref{eq:flink_factored}). For downlink,
\begin{equation}
\bar R_{\rm dl} \;\approx\; N_{\rm term}\,\eta_{\rm link}\, f_{\rm vis}\, p_{\rm avail}\, f_{\rm sched}\, R_{\rm pk}.
\label{eq:Ravg_model}
\end{equation}
The implied daily delivered volume is
\begin{equation}
V_{\rm day}\;\approx\;10.8\ \mathrm{TB\,day^{-1}}\Big(\frac{\bar R_{\rm dl}}{1~\mathrm{Gb\,s^{-1}}}\Big),
\label{eq:Vday}
\end{equation}
using decimal TB.

For an aggressive but representative \(f_{\rm vis}=0.2\), \(\eta_{\rm link}=0.7\), \(f_{\rm sched}=1\), and \(N_{\rm term}=1\), Eq.~(\ref{eq:Ravg_model}) gives \(\bar R_{\rm dl}\approx 7.0\), \(13.1\), and \(13.9~\mathrm{Gb\,s^{-1}}\) at \(R_{\rm pk}=100~\mathrm{Gb\,s^{-1}}\) for \(N_{\rm gs}=1,4,8\) with \(p_{\rm cs}=0.5\); doubling \(R_{\rm pk}\) doubles these averages. Weather diversity therefore saturates quickly once \(p_{\rm avail}\rightarrow 1\), after which average throughput is set mainly by geometry and achievable \(R_{\rm pk}\). The important implication for C3 is scale coupling: per-node sustained rates of \(\mathcal{O}(10)\text{--}\mathcal{O}(30)~\mathrm{Gb\,s^{-1}}\) correspond to \(\mathcal{O}(0.1)\text{--}\mathcal{O}(0.3)~\mathrm{PB/day/node}\), and therefore drive both space-to-ground link infrastructure and terrestrial backhaul.

A numeric ceiling on \(\Gamma_{\max}\) for representative terminal and ground assumptions is given in Table~\ref{tab:Gamma_ceiling}.

\subsection{A communications-feasibility bound in terms of \texorpdfstring{$\Gamma$}{Gamma}}
\label{sec:Gamma_bound}

The communication-intensity definition in Sec.~\ref{sec:system} and the duty-cycle bound in Eq.~(\ref{eq:Ravg})
imply a hard feasibility ceiling for any terrestrial-user workload:
\begin{equation}
\Gamma \;\le\; \Gamma_{\max}
\equiv \frac{N_{\rm term}\,\eta_{\rm link}\, f_{\rm vis}\, p_{\rm avail}\, f_{\rm sched}\, R_{\rm pk}}{\Util\,\Pit}.
\label{eq:Gamma_max}
\end{equation}

Eq.~(\ref{eq:Gamma_max}) converts the communications subsystem into an explicit workload filter. For fixed \(N_{\rm term}\), \(R_{\rm pk}\), and duty factor, the ceiling scales as \(\Gamma_{\max}\propto \Pit^{-1}\); increasing node power from \(\SI{1}{MW}\) to \(\SI{10}{MW}\) without proportional communications scaling therefore reduces admissible I/O-per-energy by an order of magnitude. Preserving a fixed \(\Gamma_{\max}\) at larger \(\Pit\) requires proportional increases in terminal count, peak rate, or usable contact fraction, all of which raise terminal mass or electrical overhead and feed back into \(\alphaOH\) and \mkw[{\rm fixed}]. The same ceiling in data-per-energy units is
\begin{equation}
\Big(\frac{D}{E_{\rm IT}}\Big)_{\rm kWh}
\le 3.6\times 10^6\,\Gamma_{\max}
\qquad [\mathrm{bit/kWh}_{\rm IT}].
\label{eq:Gamma_kWh}
\end{equation}
Using Table~\ref{tab:Gamma_ceiling}, the representative \(R_{\rm pk}=200~\mathrm{Gb\,s^{-1}}\) configuration yields \(\Gamma_{\max}\approx 3.28\times10^4~\mathrm{bit/J}\) at \(\Pit=\SI{1}{MW}\), corresponding to \(14.8~\mathrm{GB/kWh}_{\rm IT}\), but only \(1.48~\mathrm{GB/kWh}_{\rm IT}\) at \(\Pit=\SI{10}{MW}\). In other words, high-\(\Gamma\), storage-coupled, or strongly interactive services are excluded by architecture before capital cost is even considered.

\begin{table}[t]
\caption{Illustrative communication-intensity ceiling $\Gamma_{\max}$ and the corresponding
data-per-energy ceiling in GB/kWh$_{\rm IT}$ for a representative comm configuration ($R_{\rm pk}=200$~Gb/s, $f_{\rm vis}=0.2$,
$N_{\rm gs}=4$, $p_{\rm cs}=0.5$, $\eta_{\rm link}=0.7$, $f_{\rm sched}=1$, $N_{\rm term}=1$, $\Util=0.8$).}
\label{tab:Gamma_ceiling}
\centering
\small
\setlength{\tabcolsep}{6pt}
\renewcommand{\arraystretch}{1.12}
\begin{tabular}{c c c}
\toprule
$\Pit$ (MW) & $\Gamma_{\max}$ (bit/J) & Data ceiling (GB/kWh$_{\rm IT}$) \\
\midrule
0.1 & $3.28\times 10^{5}$ & 148 \\
1   & $3.28\times 10^{4}$ & 14.8 \\
10  & $3.28\times 10^{3}$ & 1.48 \\
\bottomrule
\end{tabular}
\end{table}

Workloads that are intrinsically high-$\Gamma$ (large bidirectional I/O tightly coupled to terrestrial storage or users) are therefore structurally disfavored in orbit, while low-$\Gamma$ workloads (compute-dense, delay-tolerant jobs; training with pre-positioned datasets; space-native preprocessing) can fit beneath the ceiling even if $\bar R_{\rm dl}$ is only $\mathcal{O}(10)$~Gb/s per node. For training-style workloads, low effective \(\Gamma\) usually requires dataset pre-positioning; otherwise the initial uplink campaign can dominate operational feasibility.

\subsection{Bidirectional asymmetry and the space-to-ground portion of traffic}
\label{sec:comms_asym}

The scalar \(\Gamma\) provides a first-order workload filter, but real workloads are often asymmetric. Dataset staging,
checkpoints, control traffic, model updates, and result downlink do not necessarily scale together. A more explicit
representation is
\begin{equation}
\Gamma_{\uparrow} \equiv \frac{D_{\uparrow}}{E_{\rm IT}}, \qquad
\Gamma_{\downarrow} \equiv \frac{D_{\downarrow}}{E_{\rm IT}}, \qquad
\Gamma = \Gamma_{\uparrow}+\Gamma_{\downarrow},
\label{eq:Gamma_ud}
\end{equation}
with separate feasibility ceilings
\begin{equation}
\Gamma_{\uparrow,\max} \equiv \frac{N_{{\rm term},\uparrow}\,\eta_{{\rm link},\uparrow}\,f_{{\rm vis},\uparrow}\,p_{{\rm avail},\uparrow}\,f_{{\rm sched},\uparrow}\,R_{{\rm pk},\uparrow}}{\Util\,\Pit},
\qquad
\Gamma_{\downarrow,\max} \equiv \frac{N_{{\rm term},\downarrow}\,\eta_{{\rm link},\downarrow}\,f_{{\rm vis},\downarrow}\,p_{{\rm avail},\downarrow}\,f_{{\rm sched},\downarrow}\,R_{{\rm pk},\downarrow}}{\Util\,\Pit}.
\label{eq:Gamma_ud_max}
\end{equation}

The main-text scalar ceiling in Eq.~\ref{eq:Gamma_max} should therefore be read as a compact summary of a generally
asymmetric pair of constraints. This matters operationally: a workload may be downlink-feasible but uplink-limited
if large training datasets must be refreshed frequently from Earth, or uplink-feasible but downlink-limited if result
products are bulky. It also matters economically because only the space-to-ground fraction of traffic, \(\Gamma_{\rm ext}\)
from Eq.~\ref{eq:Gamma_ext_fraction}, should be charged to the space-to-ground communications term \(\CGS\).

\section{Radiation, fault tolerance, and delivered compute-years}
\label{sec:radiation}

Radiation is a first-order systems constraint for orbital compute because it impacts both (i) \emph{short-timescale performance} via single-event effects (SEEs) and (ii) \emph{lifetime economics} via total ionizing dose (TID), displacement damage, and the probability of unrecoverable failure \cite{Felix2024JetsonOrinTID,Rodriguez2025JetsonSEE}. These effects are strongly orbit-dependent (altitude, inclination, South Atlantic Anomaly exposure, shielding, and solar cycle), so this section remains parametric and focuses on how the radiation environment enters \(\Util\), lifetime hazard \(\lambda\), and ultimately cost per delivered compute-year \cite{AE9AP9}.  

Orbital compute economics depend on \emph{delivered compute-years} rather than nameplate compute.
Radiation enters twice: it drives (i) transient faults that reduce effective utilization through checkpointing/ECC overhead, and (ii) cumulative degradation and hazard rate that limits lifetime and increases replacement cadence.

\subsection{Orbit scenarios: radiation environment inputs and penalty outputs}
\label{sec:rad_scenarios}

\begin{table*}[t]
\caption{Representative orbit scenarios and their coupling into eclipse, radiation, and lifecycle penalties. Eclipse maxima are computed at \(\beta=0\). The TID values are illustrative 5-year mission anchors for silicon behind a few-mm Al-equivalent shield; they are order-of-magnitude design values rather than direct AE9/AP9 outputs. A mission-grade specification should combine AE9/AP9-IRENE trapped-particle runs with a solar proton model and an effects code such as SHIELDOSE-2.}
\label{tab:orbit_rad_cases}
\centering
\small
\setlength{\tabcolsep}{6pt}
\renewcommand{\arraystretch}{1.10}
\begin{tabular}{l c c c c c}
\toprule
Case & Orbit (illustrative) & $T_{\rm orb}$ (h) & $t_{\rm ecl,max}$ (min) & $f_{\odot,\rm min}$ & TID(5yr) [krad(Si)] \\
\midrule
LEO baseline & 550~km, mid-incl. & 1.59 & 35.6 & 0.628 & 3--6 \\
Dawn--dusk LEO & 600~km SSO (DD)  & 1.61 & $\approx 0$ & $\approx 1$ & $\approx 5$ \\
Higher LEO & 1200~km, high-incl. & 1.82 & 34.8 & 0.682 & 12--20 \\
GEO & 35,786~km GEO & 23.93 & 69.4 (seasonal) & $\sim 0.952$ & 20--30 \\
\bottomrule
\end{tabular}
\end{table*}

To close the loop from orbit selection to delivered compute-years, we define a small set of representative orbit scenarios and propagate each scenario into (i) eclipse duty parameters ($\etaSun$, $t_{\rm ecl}$), (ii) radiation environment metrics (TID and SEE driver spectra), and (iii) economic penalties through $\Util_{\rm eff}$ and $\Pi_{\rm life}$.

Radiation environments should be parameterized with AE9/AP9-IRENE for trapped particles \cite{AE9AP9}. For full-mission TID or DDD, however, trapped-particle results should be combined with a solar proton model and then folded througth an effects code such as SHIELDOSE-2 \cite{Guild2022BestPractices}. In the absence of a project-specific IRENE run, Table~\ref{tab:orbit_rad_cases} therefore uses literature-anchored 5-year TID brackets for silicon behind a few-mm Al-equivalent shield, drawing on a 600 km SSO OMERE result of \(\sim 5\) krad(Si) for 3.5 mm Al, SPENVIS dose-depth results for 600/1200 km LEO cases, and a GEO 5-year dose-depth example of \(\sim 20\) krad(Si) behind 200 mil Al \cite{Rodeck2025Cube1G,Buchner2008RHA}. The dose values in Table~\ref{tab:orbit_rad_cases} are therefore surrogate design brackets rather than percentile-specific AE9/AP9 mission outputs; a mission study would replace them with AE9/AP9-IRENE runs phased across the solar cycle and combined with solar-particle transport, shielding transport, and drag-driven replenishment analysis.

The LEO and GEO entries are intentionally given as design brackets rather than precise predictions, because shielding geometry, percentile choice, and inclusion of solar-particle contributions can materially shift the resulting dose \cite{Guild2022BestPractices,Buchner2008RHA}.

\subsection{Transient faults and utilization penalty}
\label{sec:see_util}

Let \(R_{\rm SEE}\) be the node-level rate of \emph{correctable} SEE-driven events (bit flips, link upsets, recoverable logic errors).
A generic device-physics expression is
\begin{equation}
R_{\mathrm{SEE}} \sim \int \Phi(E)\,\sigma_{\mathrm{SEE}}(E)\,dE,
\label{eq:RSEE}
\end{equation}
where \(\Phi(E)\) is the differential particle flux (orbit- and shielding-dependent) and \(\sigma_{\rm SEE}(E)\) is the effective system cross-section after architectural masking (error-correcting code (ECC) strength, interleaving, retry policies, etc.).

To connect SEE rate to \emph{delivered throughput}, separate three overhead channels \cite{AE9AP9}:
(i) proactive fault-tolerance overhead (checkpointing, redundancy),
(ii) reactive recovery overhead per event, and
(iii) re-execution lost work when an event forces rollback.

Let \(\tau_{\rm cp}\) be the checkpoint interval and \(t_{\rm cp}\) the time cost of checkpointing (including I/O and synchronization).
Let \(t_r\) be the mean recovery time per correctable event (retry/ECC correction/micro-restart).
If recoverable events require rollback to the last checkpoint, the expected lost work per event is \(\approx \tau_{\rm cp}/2\).
In the regime where the fractional losses are small, a compact utilization model is
\begin{equation}
\Util_{\rm eff}
\;\approx\;
\Util_0\Big[
1
-\frac{t_{\rm cp}}{\tau_{\rm cp}}
- R_{\rm SEE}\Big(t_r+\frac{\tau_{\rm cp}}{2}\Big)
\Big],
\label{eq:Util_eff}
\end{equation}
valid when \(t_{\rm cp}/\tau_{\rm cp}\ll 1\) and \(R_{\rm SEE}(t_r+\tau_{\rm cp}/2)\ll 1\). Eq.~(\ref{eq:Util_eff}) makes the systems point explicit:
the relevant question is whether the compound quantity \(R_{\rm SEE}(t_r+\tau_{\rm cp}/2)\) materially reduces \(\Util\), thereby increasing levelized cost in Eq.~(\ref{eq:LCIT}).

Two further couplings are worth making explicit:
(i) pushing to higher \(\Pit\) typically increases node memory and switching fabric scale, which increases the effective system cross-section \(\sigma_{\rm SEE}\) unless mitigated by stronger ECC/partitioning;
(ii) pushing to higher \(\Trad\) (Sec.~\ref{sec:thermal}) can tighten timing margins and increase the performance impact of fault recovery if throttling interacts with checkpoint cadence.

\noindent\emph{From orbit radiation spectra to $R_{\rm SEE}$ and $\Util_{\rm eff}$.}
\label{sec:rad_to_util}

For a specified orbit scenario (Table~\ref{tab:orbit_rad_cases}), $R_{\rm SEE}$ can be evaluated by combining
(i) environment spectra (LET spectra for heavy ions; proton energy spectra for trapped/solar protons),
(ii) shielding transport, and (iii) device-level SEE response functions.

A practical system-level representation for memory-dominated upsets is
\begin{equation}
R_{\rm SEU} \;\approx\; N_{\rm bit}\int \Phi_{\rm eff}(L)\,\sigma_{\rm SEU}(L)\,dL,
\label{eq:RSEU}
\end{equation}
where $N_{\rm bit}$ is the protected bit count and $\Phi_{\rm eff}(L)$ is the post-shield LET spectrum.
The node-level $R_{\rm SEE}$ in Eq.~(\ref{eq:Util_eff}) may then be formed as
\begin{equation}
R_{\rm SEE} \;=\; R_{\rm SEU} + R_{\rm SEL} + R_{\rm func},
\label{eq:RSEE_sum}
\end{equation}
including latchup (SEL) and functional interrupts (SEFI) as applicable. This explicitly separates
\emph{correctable} events (driving $\Util_{\rm eff}$ through checkpoint/recovery overhead) from
\emph{catastrophic} events, which instead contribute to the hazard rate $\lambda$ below.

\subsection{Hazard rate, lifetime, and expected delivered IT-energy}
\label{sec:lifetime_hazard}

Let \(\lambda\) be the effective hazard rate for \emph{permanent} node loss (unrecoverable SEE/latchup, TID wearout,
debris collision, propulsion failure, or forced disposal) \cite{NASAORDEM}. We intentionally aggregate failure modes into \(\lambda\) because
the economic mapping depends only on the survival function. With survival probability \(S(t)=\exp(-\lambda t)\),
the expected delivered IT-energy over mission duration \(T\) is
\begin{equation}
E_{\mathrm{IT,del}}
=\int_0^T \Util_{\rm eff}\,\Pit\,S(t)\,dt
=\Util_{\rm eff}\,\Pit\,\frac{1-e^{-\lambda T}}{\lambda}.
\label{eq:delivered_energy}
\end{equation}

Define the lifetime penalty factor
\begin{equation}
\Pi_{\rm life}\equiv \frac{\lambda T}{1-e^{-\lambda T}}\ge 1,
\label{eq:Pi_life}
\end{equation}
which multiplies the effective cost per delivered compute-year relative to an ``immortal'' node.
Two limiting cases bound the delivered-compute penalty:
\(\Pi_{\rm life}\approx 1+\lambda T/2\) for \(\lambda T\ll 1\), and \(\Pi_{\rm life}\approx \lambda T\) for \(\lambda T\gg 1\).

For reference, \(\lambda=0.10~\mathrm{yr^{-1}}\) and \(T=5~\mathrm{yr}\) give \(\Pi_{\rm life}\approx 1.27\), while \(\lambda=0.20~\mathrm{yr^{-1}}\) and the same lifetime give \(\Pi_{\rm life}\approx 1.58\).
\noindent\emph{Orbit-linked hazard rate and $\Pi_{\rm life}$.}
\label{sec:lambda_orbit}

We model catastrophic loss as a superposition of approximately independent hazards:
\begin{equation}
\lambda \;=\; \lambda_{\rm rad} + \lambda_{\rm MMOD} + \lambda_{\rm prop} + \lambda_{\rm sw} + \lambda_{\rm ops},
\label{eq:lambda_sum}
\end{equation}
where $\lambda_{\rm rad}$ captures unrecoverable radiation events and dose-driven wearout, $\lambda_{\rm MMOD}$ captures
debris/MMOD catastrophic failure, and the remaining terms capture propulsion, software, and operational hazards.

For a given orbit scenario, $\lambda_{\rm rad}$ can be parameterized in terms of (i) dose margin to end-of-life,
(ii) SEL/SEFI susceptibility for the chosen compute class, and (iii) recovery doctrine (safe-mode, reboot, scrubbing). The resulting $\lambda$ enters delivered compute-years through $\Pi_{\rm life}$ [Eq.~(\ref{eq:Pi_life})] and $E_{\rm IT,del}$ [Eq.~(\ref{eq:delivered_energy})], making orbit selection a first-order economic decision rather than a secondary implementation detail.

\subsection{Technology obsolescence and refresh lag}
\label{sec:obsolescence}

Hazard-limited lifetime is not the only delivered-compute penalty for orbital hardware. The compute payload is frozen
at manufacture and then exposed to integration, launch, commissioning, and replacement delays. If terrestrial frontier
performance per watt improves at an effective rate \(g_{\rm obs}\) [yr\(^{-1}\)], then a node that survives physically can
still lose economic competitiveness through refresh lag. A compact frontier-relative delivered-energy metric is
\begin{equation}
E_{\rm IT,del}^{\star}
=
\int_0^T U_{\rm eff}\,\Pit\,e^{-(\lambda+g_{\rm obs})t}\,dt
=
U_{\rm eff}\,\Pit\,\frac{1-e^{-(\lambda+g_{\rm obs})T}}{\lambda+g_{\rm obs}}.
\label{eq:EITdel_obs}
\end{equation}
The corresponding combined lifetime-plus-obsolescence penalty is
\begin{equation}
\Pi_{\rm life+obs}
\equiv
\frac{(\lambda+g_{\rm obs})T}{1-e^{-(\lambda+g_{\rm obs})T}}.
\label{eq:Pilife_obs}
\end{equation}

This extension preserves it preserves the same delivered-compute-year logic used for \(\Pi_{\rm life}\) while
making technology refresh explicit. For example, \(g_{\rm obs}=0.20~\mathrm{yr^{-1}}\), \(\lambda=0.10~\mathrm{yr^{-1}}\), and
\(T=5~\mathrm{yr}\) imply \(\Pi_{\rm life+obs}\approx 1.93\), compared to \(\Pi_{\rm life}\approx 1.27\) when obsolescence is
ignored. The benchmark results below retain \(\Pi_{\rm life}\) as the primary structural penalty, but project-specific cases---
especially accelerator-heavy nodes---should include \(g_{\rm obs}\) explicitly.

\subsection{Implication for break-even economics}
\label{sec:radiation_econ}

Eqs.~(\ref{eq:Util_eff})--(\ref{eq:Pi_life}) enter the economics through \emph{delivered} IT-energy.
A convenient way to express this is to take the levelized infrastructure cost per delivered IT-energy
in Eq.~(\ref{eq:LCIT}) and apply two substitutions:
\(\Util \rightarrow \Util_{\rm eff}\) (fault-tolerance and recovery overhead), and a multiplicative penalty \(\Pi_{\rm life}\)
(replacement cadence and lifetime hazard).
Thus the effective levelized infrastructure cost becomes
\begin{equation}
c_{\mathrm{orbit,IT}}^{(\mathrm{del})}
\simeq
\Pi_{\rm life}\,(1+\lambda\tau_{\rm rep})
\frac{\CRF\,C_{\mathrm{orbit}} + O_{\mathrm{orbit}}}{8760\,U_{\rm eff}}
\qquad [\$/\mathrm{kWh}_{\rm IT}].
\label{eq:LCIT_delivered}
\end{equation}
This makes the economic fragility mechanism explicit: even modest reductions in \(\Util_{\rm eff}\) and moderate
\(\lambda T\) can dominate the comparison when \(C_{\rm orbit}\) is large. It also clarifies why many C1/C2 workloads are structurally more favorable than always-on interactive C3 services: C1/C2 can tolerate duty-cycle operation and may justify higher \(\Pi_{\rm life}\) and lower \(\Util_{\rm eff}\) because location creates intrinsic value, whereas C3 must satisfy Eq.~(\ref{eq:breakeven}) \emph{and} maintain high delivered compute-years.

Eq.~(\ref{eq:LCIT_delivered}) converts reliability into economics. The factor \(\Pi_{\rm life}\) penalizes finite lifetime and replacement cadence, while \(U_{\rm eff}\) penalizes transient faults, recovery overhead, and idle time. For illustration, \(\Pi_{\rm life}=0.5/(1-e^{-0.5})\approx 1.27\) for \(\lambda=0.10~\mathrm{yr^{-1}}\) and \(T=5~\mathrm{yr}\), while \(\Pi_{\rm life}=1/(1-e^{-1})\approx 1.58\) for \(\lambda=0.20~\mathrm{yr^{-1}}\) and the same lifetime. Thus a node that is close to capital-cost break-even can be pushed \(27\%\)–\(58\%\) higher in delivered-cost terms before any additional reduction in \(U_{\rm eff}\) is included. Orbit choice, fault doctrine, and replenishment latency are therefore first-order economic variables, not secondary implementation details.

\subsection{Correlated fault domains and architecture resilience}
\label{sec:security}

The monolith-versus-constellation trade is economically relevant because correlated failures raise delivered-cost even when independent node hazard remains modest. Smaller nodes reduce the capacity fraction removed by a single loss event, but they also duplicate fixed mass and increase operational complexity. The quantity that matters is therefore not node count alone, but the size of the correlated-loss domain.

We model correlated loss through an effective hazard
\begin{equation}
\lambda_{\rm eff}=\lambda+\phi\,\lambda_{\rm dom},
\label{eq:lambda_eff}
\end{equation}
where \(\lambda\) is the independent node-loss hazard, \(\lambda_{\rm dom}\) is the hazard of a domain-level event, and \(\phi\) is the capacity fraction removed when such an event occurs. This substitution is sufficient for the benchmark economics because \(\lambda_{\rm eff}\) feeds directly into \(\Pi_{\rm life}\) and delivered compute-years. In the present model, update doctrine, key hierarchy, reserve policy, and rollback strategy therefore enter economically through three channels only: \(\mkwfixed\), \(C_{\rm ops}\), and \(\lambda_{\rm eff}\).

\section{Economics: explicit break-even boundaries}
\label{sec:econ}

The physics scalings in Secs.~\ref{sec:power}--\ref{sec:mkw} collapse orbital feasibility into a small number of cost-relevant parameters. Here we express these constraints in two complementary units:
(i) infrastructure capital cost per delivered IT kW (\$/kW$_{\rm IT}$), which sets the \emph{speed-to-capacity} economics, and (ii) levelized infrastructure cost per delivered IT-energy (\$/kWh$_{\rm IT}$), which exposes penalties from utilization, lifetime, and operations (Sec.~\ref{sec:radiation}).

Throughout, \(\Cterr\) denotes terrestrial \emph{facility and electrical infrastructure} cost per kW of \emph{critical IT load} (excluding accelerator/server capital cost), consistent with industry benchmarking that quotes \$/MW for critical load.  \CGS\ denotes the allocated \emph{space-to-ground link} term per delivered $kW_{\mathrm{IT}}$, whether realized through dedicated optical/RF ground infrastructure or through purchased relay service. These are distinct cost centers and are not interchangeable. On the orbital side, \(C_{\rm orbit}\) is defined per delivered IT kW at the node, so that the dominant term
\((\Lkg+\Bkg)\mkw\) is dimensionally exact: \([\$/\mathrm{kg}]\times[\mathrm{kg}/\mathrm{kW}]=[\$/\mathrm{kW}]\). The economically relevant initial-technology interpretation is C1/C2; C3 is retained as the scaling anchor for terrestrial-user competition.
first-order benchmark
\begin{equation}
C_{\rm orbit}^{\rm bench}\simeq (\Lkg+\Bkg)\mkw+\CGS+C_{\rm ops},
\label{eq:corbit_bench}
\end{equation}
which is used throughout as the primary competitiveness discriminator. The effective-cost form
\begin{equation}
C_{\rm orbit}^{\rm eff}\simeq (1+s_{\rm spare})\big(L^{\rm eff}_{\$/kg}+B^{\rm eff}_{\$/kg}\big)\mkw+\CGS+C_{\rm ops},
\label{eq:corbit_eff_interp}
\end{equation}
is retained as an extended interpretation that makes reserve fraction, delivery overhead, and industrial
cost allocation explicit. In other words, Eq.~(\ref{eq:corbit_bench}) is the paper's benchmark statement, while Eq.~(\ref{eq:corbit_eff_interp}) is a refinement used to interpret how that benchmark would shift
under more detailed logistics and manufacturing assumptions.
Selected bottom-up cost-model details are retained in the optional supplemental material.

\subsection{Terrestrial baseline}
\label{sec:ground_baseline}

A meaningful comparison depends on whether one compares the full artificial-intelligence stack (chips + facility) or only the
infrastructure portion that orbit can plausibly substitute (power delivery, cooling/heat rejection, buildings, and
site electrical). Industry cost guides report wide ranges, but artificial-intelligence-ready facility and infrastructure costs commonly fall in
the \(\mathcal{O}(10)\)–\(\mathcal{O}(40)\)~M\$/MW regime depending on region, redundancy tier, cooling approach, and
substation/transmission scope \cite{TurnerTownsend2025Index,CushmanWakefield2025Cost,Uptime2023Cost}. We therefore
represent terrestrial infrastructure cost as
\begin{equation}
\Cterr \sim (1\text{--}4)\times 10^{4}\ \mathrm{\$\,kW^{-1}},
\label{eq:Cground}
\end{equation}
and treat break-even as a \emph{region} in parameter space rather than a benchmark threshold.

Two clarifications reduce reviewer ambiguity:
(i) Eq.~(\ref{eq:Cground}) is per kW$_{\rm IT}$ (critical load), not per kW of utility service; and
(ii) terrestrial energy price \(c_e\) enters the levelized comparison (Sec.~\ref{sec:lcit}) but does not change the
capital-cost discriminator in Eq.~(\ref{eq:breakeven}).

\subsection{First-order capital-cost benchmark and accounting scope}
\label{sec:orbit_capitalcost}

Let \(\Lkg\) denote marginal launch cost to the operational orbit (including mission-unique
delivery requirements), and let \(\Bkg\) denote the fully burdened spacecraft build, integration,
and qualification cost per delivered kilogram. In this notation, \(\Bkg\) is a fully burdened recurring-plus-allocated build term: spacecraft materials, subsystem recurring cost, assembly, integration, test, quality assurance, qualification, yield loss, and allocated non-recurring engineering may all be included if they are charged per delivered kilogram. Operator-specific learning, vertical integration, and reusable-launch economics therefore enter through \(\Lkg\), \(\Bkg\), and the effective-cost extension below rather than through a separate hidden multiplier. We use the first-order capital cost benchmark
\begin{equation}
C_{\rm orbit}[\$/\mathrm{kW}]\simeq (\Lkg+\Bkg)\mkw + \CGS +C_{\rm ops},
\label{eq:corbit}
\end{equation}
as a physically interpretable discriminator between orbital and terrestrial infrastructure. Here \(\CGS\) denotes the space-to-ground communications and terminal term allocated per delivered \(\mathrm{kW}_{\rm IT}\) served, and \(C_{\rm ops}\) captures capital-cost-equivalent lifecycle costs that are not naturally expressed as kg-scaled hardware (e.g., insurance premiums capitalized over life, compliance/disposal reserves, and risk and compliance overhead that scales with node count) \cite{NASASTD8719_14B,ISO24113_2019,Turyshev_Debris_2026}. Eq.~(\ref{eq:corbit}) is not a point forecast of future industrial cost; it is a necessary capital-cost closure condition that shows how mass-per-power, launch and spacecraft build burden, space-to-ground communications cost, and operations terms jointly shape competitiveness.

Eq.~(\ref{eq:corbit}) immediately gives the capital-cost sensitivity
\[
\frac{\partial C_{\rm orbit}}{\partial(\Lkg+\Bkg)} = \mkw.
\]
Hence, if \(\mkw \simeq 40~\mathrm{kg/kW}\), every \(100~\mathrm{\$\,kg^{-1}}\) reduction in combined launch and spacecraft build burden lowers orbital capital cost by \(4\times10^3~\mathrm{\$\,kW^{-1}}\). This is why \(\mkw\) is the central economic variable: once the platform resides in the tens-of-kg/kW regime, modest changes in \((L+B)\$/kg\) correspond to multi-k\$/kW shifts in competitiveness.

We separate terrestrial-side costs into two \emph{non-overlapping} components with the same units [\$/kW$_{\rm IT}$] but different physical meaning: (i) \(C_{\rm terr}\), the levelized cost of providing IT power and facilities on Earth (electricity, cooling, buildings, and site infrastructure), and (ii) \(\CGS\), the levelized cost of the space-to-ground link required to close the space-to-ground interface (optical/RF terminals, sites and network operations, or purchased relay service) at the required availability. Accordingly, \(C_{\rm terr}\) and \(\CGS\) must not be combined unless an explicit accounting boundary is stated.
Also, if recurring operations are modeled explicitly as annual operating cost \(O_{\rm orbit}\) [\$/kW-yr] in
Eq.~(\ref{eq:LCIT}), then set \(C_{\rm ops}=0\) in Eq.~(\ref{eq:corbit}).
Conversely, if one prefers a pure capital-cost-equivalent form in Eq.~(\ref{eq:corbit}), then interpret
\(C_{\rm ops}\approx O_{\rm orbit}/\CRF\) and set \(O_{\rm orbit}=0\) in Eq.~(\ref{eq:LCIT}).
In the remainder we keep both symbols but emphasize that a consistent implementation uses \emph{one} of these two representations.

For constellation architectures, the same benchmark should be interpreted at cluster level:
\begin{equation}
C_{\rm orbit}^{(\Sigma)} \simeq (\Lkg+\Bkg)\mkw^{(\Sigma)} + \CGS^{(\Sigma)} + C_{\rm ops}^{(\Sigma)}.
\label{eq:corbit_cluster}
\end{equation}
where \(\mkw^{(\Sigma)}\) is given by Eq.~(\ref{eq:mkw_cluster}) and the space-to-ground link and operations terms are allocated over the aggregate cluster IT power \(P_{\mathrm{IT},\Sigma}\). The monolithic benchmark in Eq.~(\ref{eq:corbit}) is recovered in the limit \(N_c=1\). This paper therefore uses the monolithic case to expose the subsystem-physics floor, while interpreting competitiveness primarily through the cluster-level form in Eq.~(\ref{eq:corbit_cluster}).

\subsection{Break-even boundaries: necessary and complete forms}
\label{sec:breakeven_bounds}

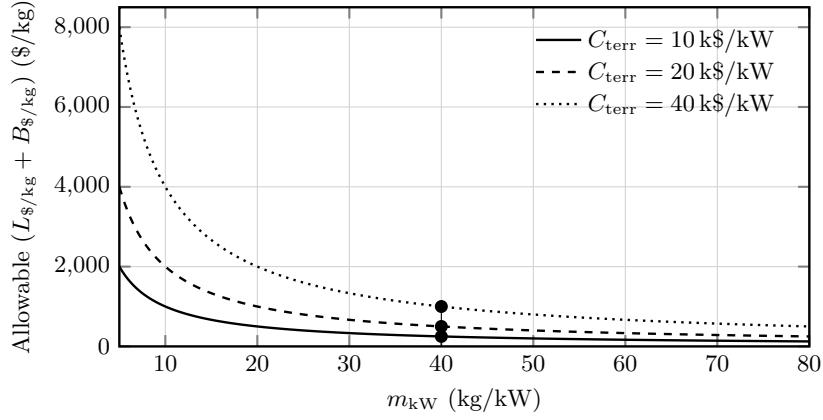
\begin{figure}[t]
\centering
\begin{tikzpicture}
\begin{axis}[
width=0.60\linewidth,
height=0.34\linewidth,
xlabel={\(\mkw\) (kg/kW)},
ylabel={Allowable \((\Lkg+\Bkg)\) (\$/kg)},
xmin=5,xmax=80,
ymin=0,ymax=8500,
grid=major,
legend style={font=\small,at={(0.78,0.97)},anchor=north,legend columns=1,draw=none,fill=none}
]
\addplot[domain=5:80,samples=250] {10000/x};
\addlegendentry{\(\Cterr=10\,\mathrm{k\$/kW}\)}

\addplot[domain=5:80,samples=250,dashed] {20000/x};
\addlegendentry{\(\Cterr=20\,\mathrm{k\$/kW}\)}

\addplot[domain=5:80,samples=250,dotted] {40000/x};
\addlegendentry{\(\Cterr=40\,\mathrm{k\$/kW}\)}

\addplot[black,thin,densely dashed,forget plot] coordinates {(40,0) (40,1100)};
\addplot[only marks,mark=*,forget plot] coordinates {(40,250) (40,500) (40,1000)};
\end{axis}
\end{tikzpicture}
\caption{Benchmark break-even boundary. The three markers at \(\mkw=40~\mathrm{kg/kW}\) correspond to the launch and spacecraft build thresholds used in the worked economic anchor. Positive \(\CGS+C_{\rm ops}\) shifts all curves downward by \((\CGS+C_{\rm ops})/\mkw\).}
\label{fig:breakeven}
\end{figure}

First suppress space-to-ground link and lifecycle terms and consider the reduced node-level proxy
\begin{equation}
(\Lkg+\Bkg)\mkw \lesssim \Cterr,
\label{eq:breakeven}
\end{equation}
which isolates the dominant physics-to-cost mapping and defines the sensitivity visualization (Fig.~\ref{fig:breakeven}). For the constellation-first architecture adopted in this paper, however, the more relevant benchmark is the cluster-level form
\begin{equation}
(\Lkg+\Bkg)\mkw^{(\Sigma)} + \CGS^{(\Sigma)} + C_{\rm ops}^{(\Sigma)} \lesssim \Cterr,
\label{eq:breakeven_full_cluster}
\end{equation}
or equivalently
\begin{equation}
\Lkg+\Bkg \lesssim \frac{\Cterr - \CGS^{(\Sigma)} - C_{\rm ops}^{(\Sigma)}}{\mkw^{(\Sigma)}}.
\label{eq:LB_bound_cluster}
\end{equation}
The node-level form is recovered in the limit \(N_c=1\), and should therefore be read as a subsystem-physics proxy rather than as the baseline deployment architecture.
For a monolithic subsystem-physics node, Eqs.~(\ref{eq:breakeven_full_cluster})--(\ref{eq:LB_bound_cluster}) reduce to the familiar node-level forms with \(\mkw^{(\Sigma)}\rightarrow\mkw\), \(\CGS^{(\Sigma)}\rightarrow \CGS\), and \(C_{\rm ops}^{(\Sigma)}\rightarrow C_{\rm ops}\).

Eqs.~(\ref{eq:breakeven})--(\ref{eq:LB_bound_cluster}) should be interpreted as benchmark thresholds rather than forecasts. Their value is to show explicitly how the required \((\Lkg+\Bkg)\) scales with \(\mkw\), \(\CGS\), and \(C_{\rm ops}\). In that sense, the economic result of this paper is structural: if \(\mkw\) remains in the tens-of-kg/kW regime, then C3 competitiveness requires unusually low launch and spacecraft build burden and unusually modest communications and lifecycle penalties.

\subsection{Ground segment and terminal cost as a throughput/availability-coupled term}
\label{sec:gs_cost}

For C3, ground segment and terminals are not optional add-ons; they scale with \emph{required average throughput}
and \emph{target availability}. A minimal parametric model consistent with Sec.~\ref{sec:comms_sanity} is
\begin{equation}
\CGS\;\approx\;
\frac{N_{\rm opt}\,C_{\rm site}+N_{\rm rf}\,C_{\rm rf}+C_{\rm backhaul}}{P_{\rm IT,name}}
\;+\;n_{\rm term}\,\frac{C_{\rm term}}{\Pit/\mathrm{kW}},
\label{eq:CGS_param}
\end{equation}
where \(C_{\rm site}\) is the capitalized cost per optical ground site (including operations and site diversity),
\(C_{\rm term}\) is the per-node optical terminal cost, \(n_{\rm term}\) is the number of terminals allocated per node (to raise \(f_{\rm link}\) and/or support multiple sites), and \(P_{\rm IT,name}\) is the aggregate nameplate IT capacity served by the ground network. 
Order-of-magnitude site capital-cost brackets of \(\sim5\text{--}50~\mathrm{M\$}\) per optical site and annual operations fractions of \(\sim5\text{--}15\%\) are adequate for the sensitivity work here; for an archival cost-scaling anchor for ground optical apertures, see Ref.~\cite{Romanofsky2019TelescopeArrayCostModel}.

To connect availability targets directly to site count, invert Eq.~(\ref{eq:pavail2}): for a desired availability
\(p_{\rm avail}^\star\),
\begin{equation}
N_{\rm gs}\ \ge\ \frac{\ln(1-p_{\rm avail}^\star)}{\ln(1-p_{\rm cs})},
\label{eq:Ng_req}
\end{equation}
under the optimistic independence assumption. Eq.~(\ref{eq:Ng_req}) provides an immediate economics lever: high \(p_{\rm avail}^\star\) drives \(N_{\rm gs}\), which drives the first term in Eq.~(\ref{eq:CGS_param}) and tightens the break-even boundary via Eq.~(\ref{eq:LB_bound_cluster}). Likewise, high-\(\Gamma\) workloads (Sec.~\ref{sec:Gamma_bound})
increase \(C_{\rm backhaul}\) and \(n_{\rm term}\), converting communications into a first-order cost driver rather than an implementation detail.
\subsection{Dedicated ground infrastructure versus purchased relay service}
\label{sec:relay_sensitivity}

The use of \(\CGS\) in Eqs.~(\ref{eq:corbit_bench})--(\ref{eq:CGS_param}) should be interpreted as a \emph{space-to-ground communications}
term, not as a commitment to dedicated optical ground sites. This sensitivity case is motivated by commercial relay-service demonstrations rather than by assuming a particular vendor architecture. NASA's Communications Services Project is funding high-rate optical relay demonstrations over commercial networks, including demonstrations using the Starlink network, while Starlink's publicly described optical capability is primarily an inter-satellite-link mesh rather than a direct optical space-to-ground service for arbitrary third-party payloads \cite{NASACommServices2026,StarlinkProgress2025}. A purchased relay-service alternative can therefore be represented parametrically by separating a terminal-integration term from a throughput-priced service term; the service price, access policy, quality of service, and end-to-end routing are left as explicit variables.

Define the space-to-ground traffic in data-per-energy form as
\begin{equation}
\Gamma_{\rm GB} \equiv 4.5\times 10^{-4}\Gamma_{\rm ext}
\qquad [\mathrm{GB/kWh}_{\mathrm{IT}}],
\label{eq:Gamma_GB}
\end{equation}
where \(\Gamma_{\rm ext}\) remains the space-to-ground communication intensity in \(\mathrm{bit/J}\), and Eq.~(\ref{eq:Gamma_GB}) is simply the corresponding unit conversion of Eq.~(\ref{eq:Gamma_GBperKWh}). The annual relay charge per delivered IT kW is then
\begin{equation}
O_{\rm relay}
=
8760\,\Util\,\Gamma_{\rm GB}\,c_{\rm GB}
\qquad [\$/(\mathrm{kW\,yr})],
\label{eq:Orelay}
\end{equation}
with \(c_{\rm GB}\) in \$/GB. The corresponding capital-cost-equivalent space-to-ground communications term is
\begin{equation}
\CGS^{(\rm relay)}
\approx
\frac{O_{\rm relay}}{\CRF}
+
N_{\rm term}\frac{C_{\rm term}}{\Pit/\mathrm{kW}},
\label{eq:CGS_relay}
\end{equation}
to be compared against the dedicated-site realization of \(\CGS\) in Eq.~(\ref{eq:CGS_param}). Purchased relay service
therefore changes the cost allocation rather than eliminating the communications bottleneck.

The relevant threshold is the relay price that makes purchased service break even
with a dedicated space-to-ground link realization:
\begin{equation}
c_{\rm GB}^{\star}
\approx
\frac{\CRF\left(\CGS^{(\rm ded)}-N_{\rm term}C_{\rm term}/(\Pit/\mathrm{kW})\right)}
{8760\,\Util\,\Gamma_{\rm GB}},
\label{eq:cGB_star}
\end{equation}
where \(\CGS^{(\rm ded)}\) is the dedicated-site realization from Eq.~(\ref{eq:CGS_param}). For example, if
\(\CGS^{(\rm ded)}=10~\mathrm{k\$/kW}\), \(\CRF=0.10\), \(\Util=0.8\), and \(\Gamma_{\rm GB}=15~\mathrm{GB/kWh}_{\mathrm{IT}}\),
then Eq.~(\ref{eq:cGB_star}) gives \(c_{\rm GB}^{\star}\approx 9.5\times 10^{-3}~\$/\mathrm{GB}\), i.e. about
\(0.95\) cents/GB before terminal-integration adjustments. Relay purchase can therefore be materially better than a
dedicated optical ground build for low-\(\Gamma_{\rm ext}\) or bursty workloads, but it still leaves high-\(\Gamma\)
workloads economically constrained because space-to-ground data movement remains first-order.

\subsection{Levelized cost per delivered IT-energy}
\label{sec:lcit}

Capital cost per delivered kilowatt is not sufficient for cross-architecture comparison if utilization, lifetime, or operations differ. Using the capital recovery factor
\begin{equation}
\CRFdef,
\label{eq:CRF}
\end{equation}
and annual operating cost \(O_{\rm orbit}\) [\$/kW-yr], the levelized infrastructure cost per delivered IT-energy is
\begin{equation}
c_{\mathrm{orbit,IT}}
\simeq
\frac{\CRF\,C_{\mathrm{orbit}} + O_{\mathrm{orbit}}}{8760\,\Util}
\quad [\$/\mathrm{kWh}_{\mathrm{IT}}],
\label{eq:LCIT}
\end{equation}
with the understanding that Sec.~\ref{sec:radiation} replaces \(\Util\) by \(U_{\rm eff}\) and introduces the lifetime penalty \(\Pi_{\rm life}\). In practice orbit competes on the \emph{infrastructure increment} relative to terrestrial provision, not on eliminating accelerator/server capital cost, which is largely common-mode across the comparison. Any claimed energy-cost advantage is therefore relevant only if it survives the space-to-ground link, utilization, and lifetime penalties that determine delivered compute-years.

\subsection{Worked example: a \texorpdfstring{\(\Pit=\SI{1}{MW}\)}{1 MW} node (physics-floor and base-case design point)}
\label{sec:worked_structure}

This section revisits the same \(\Pit=\SI{1}{MW}\) anchor introduced earlier in Sec.~\ref{sec:worked_example}, but now separates the simplified physics-floor subtotal from a fuller base-case design point with explicit non-ideal factors. The intent is to keep the early worked example as the structural sizing anchor while concentrating the more complete numerical interpretation in one place. To avoid conflating a structural lower bound with a deployable design point, we separate the \(\Pit=\SI{1}{MW}\) anchor into two cases.

\emph{Physics-floor case:} a lower-bound subtotal that isolates the irreducible PV+radiator burden
under optimistic closure assumptions,
\(
t_{\rm ecl}=0,\ q_{\rm env}=0,\ \etaPMAD=\eta_b=\fPVEOL=1
\),
and no battery term. Under these conditions, 
the PV+radiator subtotal is a lower bound,
\begin{equation}
{\mkw}_{\rm floor}
=
\frac{1000\,\alphaOH}{\etaSun\,\spPV}
+
\frac{1000\,\alphaOH\,\sigRad}{\epsrad\,\etaview\,\sigSB\,\Trad^4}.
\label{eq:mkw_floor_worked}
\end{equation}
Here \(t_{\rm ecl}=0\) suppresses the battery term, while \(\etaSun\) is retained as an effective sunlit-duty factor that can include geometric or operational margin even in a high-sunlight orbit. This quantity is reported only to expose the structural physics floor; it is not the paper's
benchmark design point.

\emph{Base-case design point:} a full-equation evaluation using explicit values for
\(\etaPMAD,\eta_b,\fPVEOL,\rhoDoD,\fBatEOL\), and \(q_{\rm env}\).
Only the base case should be used for numerical cost statements. For the base case we evaluate the full end-of-life relations using explicit non-ideal factors.
Unless otherwise stated, the illustrative base-case values are
\[
\etaPMAD=\num{0.92},\quad
\eta_b=\num{0.90},\quad
\fPVEOL=\num{0.85},\quad
\rhoDoD=\num{0.80},\quad
\fBatEOL=\num{0.80},\quad
q_{\rm env}=\SI{150}{W\,m^{-2}}.
\]
These values are base-case inputs used to separate the structural physics floor from a more realistic
design point. 

\begin{table}[t]
\caption{Single \(\Pit=\SI{1}{MW}\) anchor separating a physics-floor subtotal from a base-case design point. The physics-floor case sets \(t_{\rm ecl}=0\), \(q_{\rm env}=0\), and \(\etaPMAD=\eta_b=\fPVEOL=1\). The base case uses \(\etaPMAD=0.92\), \(\eta_b=0.90\), \(\fPVEOL=0.85\), \(\rhoDoD=0.80\), \(\fBatEOL=0.80\), and \(q_{\rm env}=\SI{150}{W\,m^{-2}}\). Common inputs are \(\alphaOH=1.25\), \(\etaSun=0.95\), \(\spPV=\SI{100}{W\,kg^{-1}}\), \(\Trad=\SI{350}{K}\), \(\epsrad=0.90\), \(\etaview=0.85\), and \(\sigRad=\SI{5}{kg\,m^{-2}}\).}
\label{tab:worked_1mw_anchor}
\centering
\small
\setlength{\tabcolsep}{4pt}
\renewcommand{\arraystretch}{1.00}
\begin{tabular}{lcc}
\toprule
Quantity & Physics floor & Base case \\
\midrule
\(\Apv\) or \(\Apv^{\rm BOL}\) & \(\SI{4.4e3}{m^2}\) & \(\SI{5.64e3}{m^2}\) \\
\(M_{\rm PV}\) or \(M_{\rm PV}^{\rm BOL}\) & \(\SI{1.32e4}{kg}\) & \(\SI{1.69e4}{kg}\) \\
\(\Arad\) & \(\SI{1.92e3}{m^2}\) & \(\SI{2.50e3}{m^2}\) \\
\(M_{\rm rad}\) & \(\SI{9.6e3}{kg}\) & \(\SI{1.25e4}{kg}\) \\
\(\mkwPV\) & \(13.2\) & \(16.9\) \\
\(\mkwrad\) & \(9.6\) & \(12.5\) \\
\(\mkwbat\) & \(0\) & \(0\) \\
\hline
Subtotal \(\mkwPV+\mkwrad+\mkwbat\) & \(22.8\) & \(29.4\) \\
Illustrative \(\mkwfixed\) & \multicolumn{2}{c}{\(5\text{--}30~\mathrm{kg/kW}\)} \\
Total \(\mkw\) & \(28\text{--}53\) & \(34\text{--}59\) \\
\bottomrule
\end{tabular}
\end{table}

Table~\ref{tab:worked_1mw_anchor} separates two different uses of the anchor. The physics-floor column is a lower bound that isolates the irreducible PV+radiator burden; the base-case column is the appropriate design point for economic interpretation. Relative to the floor, the base case increases PV area by \(28\%\), radiator area by \(30\%\), and the PV+radiator+battery subtotal from \(22.8\) to \(29.4~\mathrm{kg/kW}\), i.e.\ by \(29\%\). Any benchmark cost statement based on the floor therefore understates the subsystem burden by approximately one-third. The base case, not the floor, should control the abstract, conclusions, and benchmark cost discussion.

The resulting BOL PV sizing and radiator sizing are
\begin{align}
\Apv^{\rm BOL}
&=
\frac{\Ptot}{\etaPMAD\,\etaPV\,\etaGeom\,\Ssol}
\left[1+\frac{1-\etaSun}{\etaSun\,\eta_b}\right]\frac{1}{\fPVEOL},
\label{eq:Apv_base}\\
M_{\rm PV}^{\rm BOL}
&=
\frac{\Ptot}{\etaPMAD\,\spPV}
\left[1+\frac{1-\etaSun}{\etaSun\,\eta_b}\right]\frac{1}{\fPVEOL},
\label{eq:MPV_base}\\
\Arad
&=
\frac{\Ptot}{\epsrad\,\etaview\,\sigSB\,\Trad^4-q_{\rm env}},
\qquad
M_{\rm rad}=\sigRad\,\Arad.
\label{eq:Arad_base}
\end{align}

For the same high-sunlight anchor used in Sec.~\ref{sec:worked_example} (\(t_{\rm ecl}=0\), \(q_{\rm PV}=\SI{300}{W\,m^{-2}}\),
\(\alphaOH=1.25\), \(\epsilon_{\rm rad}=0.90\), \(\eta_{\rm view}=0.85\), and
\(\sigma_{\rm rad}=\SI{5}{kg\,m^{-2}}\)), Eqs.~(\ref{eq:mkw_floor_worked})–(\ref{eq:Arad_base}) give the numerical base-case values summarized in Table~\ref{tab:worked_1mw_anchor}.

\subsection{Break-even implication of the worked \texorpdfstring{\(\Pit=\SI{1}{MW}\)}{1 MW} node}
\label{sec:breakeven_1mw}

For a representative MW-class node with \(\mkw\simeq 40~\mathrm{kg/kW}\), Eq.~(\ref{eq:breakeven}) implies the benchmark condition
\begin{equation}
(\Lkg+\Bkg)\lesssim
\begin{cases}
250~\$/\mathrm{kg}, & \Cterr=10~\mathrm{k\$/kW},\\
500~\$/\mathrm{kg}, & \Cterr=20~\mathrm{k\$/kW},\\
1000~\$/\mathrm{kg}, & \Cterr=40~\mathrm{k\$/kW}.
\end{cases}
\label{eq:LB_bound_1mw}
\end{equation}
These values are benchmark thresholds; any nonzero \(\CGS\), \(C_{\rm ops}\), or lifetime penalty tightens them further.

Including space-to-ground link and lifecycle terms tightens this immediately via Eq.~(\ref{eq:LB_bound_cluster}): each additional \(1~\mathrm{k\$/kW}\) in
\(\CGS+C_{\rm ops}\) reduces the allowable \((\Lkg+\Bkg)\) by \(\approx (10^3/\mkw)\simeq 25~\$/\mathrm{kg}\) for \(\mkw=40~\mathrm{kg/kW}\). Eq.~(\ref{eq:LB_bound_1mw}) is therefore a compact, directly verifiable benchmark statement: for MW-class terrestrial-user nodes with \(\mkw\) in the tens-of-kg/kW regime, orbital competitiveness requires combined launch and spacecraft build costs in the low-\(10^2\)–low-\(10^3\)~\$/kg range and requires that \(\CGS\), \(C_{\rm ops}\), utilization, and lifetime penalties not erase the margin.

\section{Workload regimes and technology implications}
\label{sec:business_cases}

The model separates ODC applications by the mechanism through which orbit creates value. This is the practical role of the C1--C3 classification. The regimes are not alternative physics models; they are three ways of changing the terms in Eq.~(\ref{eq:C3_joint}). C1 reduces required space-to-ground data, C2 reallocates fixed spacecraft and space-to-ground communications cost across a communications mission, and C3 competes directly with terrestrial infrastructure.

\emph{C1: space-native preprocessing and autonomy.} Space-native processing is the strongest early regime because it attacks the most restrictive orbital bottleneck: the space-to-ground link. If onboard processing reduces downlink volume by a factor \(g>1\), then the effective space-to-ground communication intensity scales as \(\Gamma_{\rm ext}\rightarrow \Gamma_{\rm ext}/g\). The gain is multiplicative and acts before the communications ceiling is applied. Representative workloads include remote-sensing triage, cloud screening, region-of-interest extraction, instrument preprocessing, onboard navigation, fault management, and autonomous response. The relevant comparison is mission value per added kilogram and watt, not terrestrial facility capital cost.

\emph{C2: communications-integrated edge compute.} Communications-integrated compute is the second favorable regime because it shares the spacecraft bus, pointing system, gateway layer, and operations infrastructure with a primary communications or relay mission. In the present model this means only an allocated fraction of \(\mkwfixed\), \(\CGS\), and \(C_{\rm ops}\) is assigned to compute. Representative workloads include routing optimization, schedule optimization, edge caching, protocol translation, delay-tolerant networking functions, anomaly detection, and in-network security analytics. The governing test is incremental: whether the compute payload increases \(\alphaOH\), aperture count, gateway duty factor, or \(\Gamma_{\rm ext}\) enough to erase the shared-infrastructure advantage.

\emph{C3: terrestrial-user general compute.} terrestrial-user general compute is the scaling anchor because orbit contributes little intrinsic location value. A C3 cluster must satisfy the full joint condition: low cluster-averaged \(\mkw^{(\Sigma)}\), an admissible workload intensity \(\Gamma\le \Gamma_{\max}^{(\Sigma)}\), high \(U_{\rm eff}\), and acceptable lifetime penalty \(\Pi_{\rm life}\). The representative C3 workload set is therefore narrow: compute-dense, delay-tolerant jobs with staged datasets and low external I/O, such as Monte Carlo campaigns, parameter sweeps, offline rendering or encoding, and selected large-scale numerical simulations. Interactive services coupled tightly to terrestrial storage or users are poor candidates because they demand both high average throughput and low latency.

The regime ranking is not a qualitative preference; it follows from the equations. C1 improves feasibility by reducing \(\Gamma_{\rm ext}\). C2 improves feasibility by reducing the allocated values of \(\mkwfixed\), \(\CGS\), and \(C_{\rm ops}\). C3 has neither advantage and therefore survives only when mass, space-to-ground link, utilization, and lifetime close together. This makes the classification directly relevant to current ODC concepts: concepts that process orbital data are C1-like, concepts embedded in relay networks are C2-like, and concepts that displace terrestrial cloud or artificial-intelligence capacity are C3-like.

\section{Model scope and validation requirements}
\label{sec:limitations}

The model is a necessary-condition analysis for ODC architectures. It determines whether a proposed architecture lies in a physically and economically admissible region before detailed spacecraft design begins. A mission implementation would replace aggregate terms with a bill of materials, orbit-specific environment simulations, service-level agreements, and reliability allocations, but the governing variables remain \(\mkw\), \(\Gamma_{\max}\), \(U_{\rm eff}\), \(\Pi_{\rm life}\), \(\CGS\), and the launch and spacecraft build burden.

\emph{Compute architecture.} The compute payload is represented by \(P_{\rm IT}\), \(\alphaOH\), \(U_{\rm eff}\), and \(\Gamma\). Memory hierarchy, accelerator interconnect, checkpointing, and intra-cluster synchronization enter the spacecraft model through overhead power, effective utilization, and space-to-ground traffic. Processor-level design refines those inputs and can move an architecture in the constraint space; it does not change the form of the mass, communications, and delivered-compute inequalities.

\emph{Reference-design closure.} The fixed-mass term \(\mkwfixed\) is bounded with two reference architectures and an anchored distributed cluster. This is the appropriate level for a competitiveness-bound analysis because the photovoltaic, storage, and radiator floor is architecture-invariant, whereas structure, guidance and control, propulsion, communications, shielding, and compute packaging are architecture-sensitive. Mission-level design would narrow Tables~\ref{tab:mkw_fixed_budgets} and \ref{tab:ref_cluster_nominal}; the resulting fixed mass still adds to the subsystem floor in Eq.~(\ref{eq:mkw_decomp}).

\emph{Industrial production and vertical integration.} Launch and build burden are explicit variables. Reusable launch, vertically integrated manufacturing, yield improvement, satellite learning curves, and internal transfer prices enter through \(\Lkg\), \(\Bkg\), and the effective build model. Operator-specific economics are obtained by substituting those values into Eqs.~(\ref{eq:LB_bound_cluster})--(\ref{eq:LB_bound_1mw}).

\emph{Thermal validation.} The radiator model uses the nonlinear Stefan--Boltzmann term and the denominator penalty from absorbed environmental heat. Hardware design must resolve panel gradients, transient shadowing, loop stability, puncture isolation, deployment loads, and coolant compatibility. Those analyses act on the same three variables that control the architecture-level bound: radiator temperature \(\Trad\), absorbed environmental flux \(q_{\rm env}\), and effective radiator areal density \(\sigma_{\rm rad}\).

\emph{Space-to-ground link realization.} The symbol \(\CGS\) denotes the space-to-ground communications and terminal cost allocated per delivered IT kilowatt. It can be realized by a dedicated optical/radio-frequency ground network, purchased relay service, or a hybrid. Section~\ref{sec:relay_sensitivity} writes the relay substitution explicitly by separating terminal integration from throughput-priced service. Commercial pricing, quality-of-service guarantees, weather diversity, and gateway access determine the numerical value of \(\CGS\); the structure of the bound remains Eq.~(\ref{eq:C3_joint}).

\emph{Environment and replacement.} Radiation, drag, debris, and solar-cycle phasing determine \(\lambda_{\rm eff}\), \(\tau_{\rm rep}\), disposal reserve, and replacement cadence. A mission implementation should replace the bracketed radiation cases with percentile-specific AE9/AP9/SPM runs, solar-particle transport, shielding transport, and solar-cycle-phased drag analysis. Servicing or in-space manufacturing enters through revised \(\tau_{\rm rep}\), \(\lambda_{\rm eff}\), and \(\mkwfixed\).

\section{Conclusions}
\label{sec:conclusion}
\label{sec:concl}

Orbital data centers are technically feasible spacecraft systems, but they are not automatically competitive substitutes for terrestrial data centers. The governing issue is not access to solar energy in orbit. It is simultaneous closure of deployed mass per delivered information-technology kilowatt, radiative heat rejection, sustained space-to-ground throughput, effective utilization, lifetime, replacement, and space-to-ground communications cost. The controlling quantities are \(\mkw\), \(\Gamma_{\max}\), \(\Gamma_{\rm ext}\), \(U_{\rm eff}\), \(\Pi_{\rm life}\), and \(\CGS\). These quantities are coupled: increasing overhead power increases both photovoltaic and radiator mass; increasing space-to-ground communications capacity increases terminal power, pointing burden, and service cost; and changing node power changes fixed-mass duplication, gateway scaling, and the fraction of service lost when a node fails.

The megawatt-scale anchor gives the physical scale of the problem. For \(P_{\rm IT}=\SI{1}{MW}\) in a high-sunlight orbit, the base case requires \(\Apv^{\rm BOL}\approx\SI{5.64e3}{m^2}\), \(\Arad\approx\SI{2.50e3}{m^2}\), and \(\mkwPV+\mkwbat+\mkwrad\approx\SI{29.4}{kg/kW}\) before fixed spacecraft mass is added. With structure, deployment hardware, propulsion, communications terminals, avionics, shielding margin, and compute packaging, the total becomes \(\mkw\approx34\text{--}59~\mathrm{kg/kW}\). Thus a megawatt of orbital compute implies thousands of square meters of deployed photovoltaic and radiator area and tens of kilograms per delivered kilowatt before any terrestrial-user service is provided.

The economic implication is stringent. At \(\mkw\simeq\SI{40}{kg/kW}\), terrestrial infrastructure benchmarks of \(10\), \(20\), and \(40~\mathrm{k\$/kW}\) allow only about \(250\), \(500\), and \(1000~\mathrm{\$/kg}\), respectively, for combined launch and spacecraft build before space-to-ground link, operations, utilization, and lifetime penalties. SpaceX's published Falcon~9 service sheet lists \(\SI{74}{M\$}\) for up to \(\SI{22000}{kg}\) to low Earth orbit, corresponding to \(\SI{3.36e3}{\$/kg}\) for launch price alone under full payload utilization \cite{SpaceXCapabilities2026}; the \(250\)--\(1000~\mathrm{\$/kg}\) combined launch-plus-build allowance is therefore about a factor of 3.4--13.5 lower than that current published launch price before spacecraft build is included. Each additional \(\SI{1}{k\$/kW}\) of space-to-ground link or operations burden reduces the allowable launch and spacecraft build burden by about \(\SI{25}{\$/kg}\) at this mass-per-power level. Finite lifetime and replacement delay further multiply the delivered-cost threshold. The conclusion for C3 is therefore conditional and narrow: terrestrial-user general compute is not a generally closed replacement for terrestrial data centers under the baseline assumptions; it becomes credible only in the joint low-\(\mkw\), low-\(\Gamma_{\rm ext}\), high-\(U_{\rm eff}\), long-life regime with very low launch and spacecraft build burden. Conversely, if launch-plus-build burden remains well above the low-\(10^2\) to low-\(10^3\)~\$/kg band, or if the workload requires sustained high space-to-ground data exchange, the C3 case does not close even if the power subsystem itself is technically feasible.

The favorable regimes are different. C1 space-native processing can close because computation reduces \(D_{\rm sg}\) or enables local action at the spacecraft; the value mechanism is reduction of space-to-ground traffic and improved mission utility per kilogram and watt. C2 communications-integrated edge compute can close because compute is incremental to a relay or broadband constellation that already pays for pointing, gateway access, crosslinks, and operations. C3 has neither of those relaxations and must satisfy the full mass, communications, delivered-compute-years, and cost inequality while competing against region-specific terrestrial infrastructure. This is the technical reason the C1/C2/C3 terminology is retained: each class changes a different term in the same equations.

The technology priorities follow quantitatively from the model. Photovoltaic mass improves approximately as \(\spPV^{-1}\). Radiator area improves nominally as \(\Trad^{-4}\), but only if junction temperature, thermal transport, environmental heat absorption, puncture tolerance, and areal density close simultaneously. Communications improvements must be evaluated by sustained average space-to-ground communications capacity and allocated service cost, not by peak optical rate alone. Lifetime improvements must increase delivered compute-years, not only nominal mission duration, because utilization loss, replacement delay, obsolescence, and correlated fault domains enter the cost boundary directly.

A complete ODC architecture assessment should therefore report cluster-averaged \(\mkw\), sustained \(\Gamma_{\max}\), expected \(\Gamma_{\rm ext}\) for target workloads, \(U_{\rm eff}\), \(\Pi_{\rm life}\), space-to-ground link realization and cost, reserve/replenishment policy, and the implied launch and spacecraft build threshold. Because terrestrial cost and deliverability are regional, the economic comparison must be repeated with local \(C_{\rm terr}\), \(\chi\), electricity price, interconnection delay, and siting constraints. The result is a technology assessment rather than an advocacy argument: ODCs are credible where orbital location or infrastructure sharing changes the value equation, and they are not credible as commodity terrestrial substitutes unless the mass, communications, and delivered-lifetime inequalities close quantitatively. The framework developed here provides the technical basis for making that comparison for current and future ODC architectures.

\section*{Acknowledgments}
The work described here was carried out at the Jet Propulsion Laboratory, California Institute of Technology, Pasadena, California, under a contract with the National Aeronautics and Space Administration. 


%

\end{document}